\documentclass[
aps,prl, 
twocolumn,
reprint,
amsmath,amssymb,
superscriptaddress,
notitlepage
]
{revtex4-2}

\usepackage{graphicx}
\usepackage{amsthm}
\usepackage{cancel}
\usepackage{bbold}
\usepackage{wrapfig}
\usepackage{booktabs}
\usepackage{float}
\usepackage{enumitem}
\usepackage{hyperref}
\usepackage{tikz}
\usepackage{comment}
\usepackage{longtable}
\usepackage{multirow}
\usepackage{hyperref}
\usepackage{cleveref}

\usepackage[normalem]{ulem}

\begin{document}
\global\long\def\pgr{\mathcal{P}_{\text{gr}}}
\global\long\def\pdb{\mathcal{P}_{\text{db}}}
\global\long\def\pov{\mathcal{P}_{\text{ov}}}
\global\long\def\pn{\mathcal{P}_{0}}
\global\long\def\df{d_{\text{f}}}

\title{Explosive Transitions in Epidemic Dynamics}

\author{Georg B\"orner}
\thanks{These authors contributed equally.}
\affiliation{
  Chair for Network Dynamics,
  Center for Advancing Electronics Dresden (cfaed) and Institute for Theoretical Physics, Technische Universität Dresden, 01062 Dresden, Germany
}

\author{Malte Schr\"oder}
\thanks{These authors contributed equally.}
\affiliation{
  Chair for Network Dynamics,
  Center for Advancing Electronics Dresden (cfaed) and Institute for Theoretical Physics, Technische Universität Dresden, 01062 Dresden, Germany
}

\author{Davide Scarselli}
\affiliation{Institute of Science and Technology Austria, Klosterneuburg, Austria}

\author{Nazmi Burak Budanur}
\affiliation{Institute of Science and Technology Austria, Klosterneuburg, Austria}
\affiliation{Max Planck Institute for the Physics of Complex Systems, Dresden, Germany}

\author{Björn Hof}
\affiliation{Institute of Science and Technology Austria, Klosterneuburg, Austria}

\author{Marc Timme}
\email{Correspondence should be addressed to: marc.timme@tu-dresden.de}
\affiliation{
  Chair for Network Dynamics,
  Center for Advancing Electronics Dresden (cfaed) and Institute for Theoretical Physics, Technische Universität Dresden, 01062 Dresden, Germany
}
\affiliation{
  Cluster of Excellence Physics of Life, Technical University of Dresden
  01062 Dresden, Germany
}
\affiliation{
  Lakeside Labs, Lakeside B04b, 9020 Klagenfurt, Austria
}

\begin{abstract}
Standard epidemic models exhibit one continuous, second order phase transition to macroscopic outbreaks. However, interventions to control outbreaks may fundamentally alter epidemic dynamics. 
Here we reveal how such interventions modify the type of phase transition. In particular, we uncover three distinct types of explosive phase transitions for epidemic dynamics with capacity-limited interventions. 
Depending on the capacity limit, interventions may (i) leave the standard second order phase transition unchanged but exponentially suppress the probability of large outbreaks, (ii) induce a first-order discontinuous transition to macroscopic outbreaks, or (iii) cause a secondary explosive yet continuous third-order transition. 
These insights highlight inherent limitations in predicting and containing epidemic outbreaks. More generally our study offers a cornerstone example of a third order explosive phase transition in complex systems.
\end{abstract}

\maketitle
Phase transitions separate qualitatively different collective states emerging in large complex systems \cite{anderson1992infectious, daley2001epidemic, huang2009introduction, wolf1998spontaneous, huang2009introduction, timme2020disentangling, sethna2021statistical}. Many models of complex systems dynamics, for instance the standard susceptible-infected-recovered (SIR) model of epidemic dynamics and models of random percolation, exhibit a single  phase transition that often is second order and thus continuous \cite{kermack1927contribution, Hethcote2000mathematics, House2012ModellingEpidemicsNetworks}. For epidemic spreading dynamics, a continuous transition implies that the total number of individuals infected during an epidemic continuously varies with the infectiousness. 

Previous research has shown that complex systems may exhibit more intricate and involved collective dynamics and include discontinuous or explosive transitions if the settings become strongly nonlinear, severely constrained or heterogeneous. Examples include a strong dependence of the epidemic transition on the connectivity in structured populations with scale-free interaction topology \cite{eguiluz2002_epidemic, nielsen2020_superspreading}, epidemics where treatment options are limited by resource availability \cite{bottcher2015diseaseInducedResourceConstraints} and discontinuous hybrid phase transitions of co-evolving epidemics of two or more diseases \cite{cai2015avalanche, grassberger2016_coninfections}. Recent related results for explosive percolation processes, however, indicate that such explosive transitions might only appear discontinuous in finite size systems but are often continuous with non-standard critical exponents \cite{achlioptas2009explosive_original, nagler2011impact, costa2010explosive, riordan2011explosive, grassberger2011continuous, schroder2013crackling, dSouza2015anomalousExplosivePercolation_review, dSouza2019explosive}. 
To the best of our knowledge, all phase transitions reported to date, across epidemic models, are standard continuous, second order, or standard discontinuous, first order transitions. In this Letter, we demonstrate that capacity-limited interventions may induce explosive transitions that may appear discontinuous but in fact are third order. 

The COVID-19 pandemic has highlighted the importance of interventions such as testing, contact tracing and vaccinations to control the spread of epidemics \cite{scarselli2021discontinuous}. Importantly, the limited capacity of such interventions restricts their capability to contain epidemic outbreaks, especially when the time scales of test or vaccination rates are similar to those of the disease spread and progression. Recent empirical observations and modeling studies \cite{vyska2016complexEpidemicsWithControl, bottcher2015diseaseInducedResourceConstraints, diMuro2018multipleOutbreaksWithVaccination, scarselli2021discontinuous} suggest that interventions may prevent outbreaks or reduce their size, yet total case numbers may rapidly increase once the intervention capacity limit is reached, leading to sudden explosive and apparently discontinuous transitions to large outbreaks. 
However, the exact type of phase transitions and the mechanisms underlying them remain unclear. 

Here, we uncover explosive phase transitions emerging in epidemic models with limited-capacity interventions. We find three distinct types of transitions depending on the scaling of the intervention capacity with the total population size. We clarify the mechanisms underlying these transitions by providing generic arguments under which conditions these transitions emerge, valid for a broad class of models. More generally, our results highlight an example of a third order explosive phase transition in a generic complex system.

\begin{figure*}[ht]
    \centering
    \includegraphics[width=1.0\textwidth]{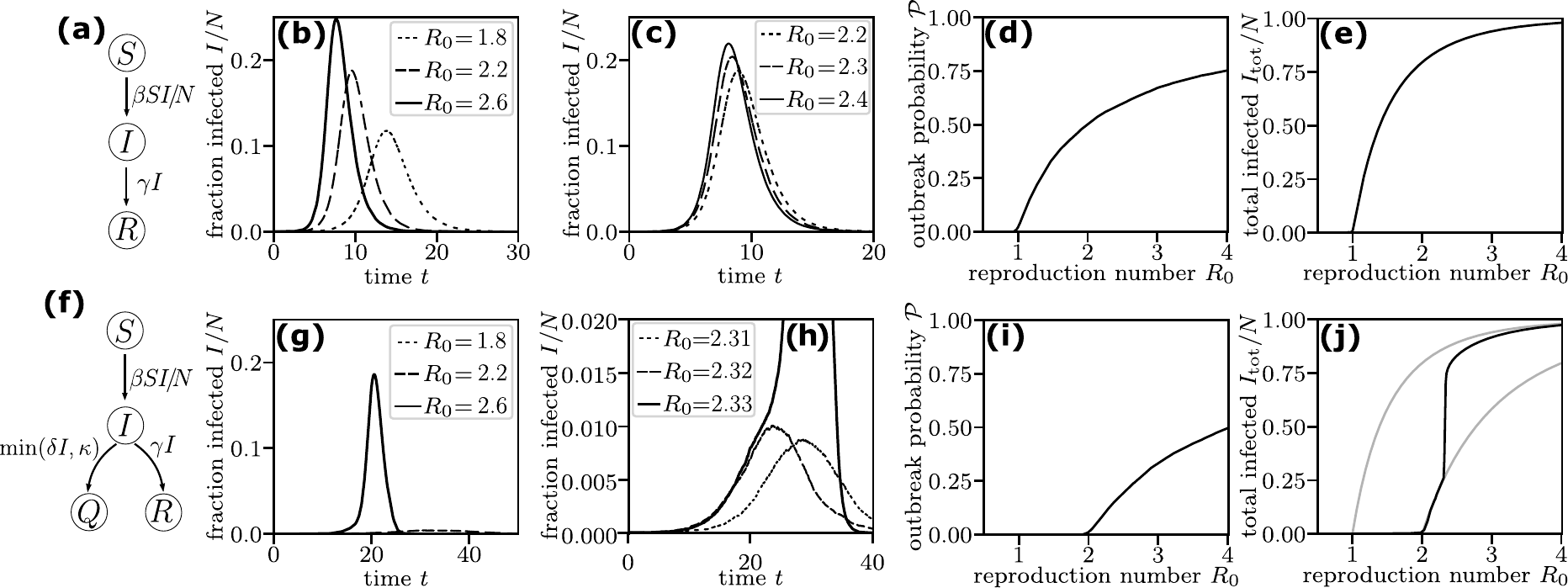}
    \caption{\textbf{Impact of capacity-limited interventions on epidemic spreading processes.} 
    Dynamics of the standard SIR model (a-e) and a model with capacity-limited interventions (SIRQ model, panels f-j). 
    (a,f) Compartment model sketch of both models. 
    (b-c,g-h) Dynamics of the number of currently infected $I(t)$ during a typical outbreak for three reproduction numbers $R_0$, each starting from $I_1(0) = 10$ initially infected. Outbreak dynamics change smoothly in the SIR model (b,c) but vary much more strongly under the influence of interventions (g,h). 
    (d,i) The outbreak probability $\mathcal{P}$ quantified by the fraction of realizations with a large number $I_\mathrm{tot} > \sqrt{N}$ of total infected out of $10^4$ realizations, each with $I_1(0) = 1$. Without interventions, large outbreaks emerge already for $R_0 > R_\mathrm{c}^{(1)} = 1$ (panel d); interventions strongly suppress the outbreak probability as long as $R_0 < R_\mathrm{c}^{(2)} = 2$ (panel i). 
    (e,j) The average total size of outbreaks grows continuously from zero once $R_0 > R_\mathrm{c}^{(1)}$ for the SIR model without interventions or $R_0 > R_\mathrm{c}^{(2)}$ for the SIRQ model, respectively. With limited intervention capacity, a secondary transition emerges at $R_\mathrm{c}^{(3)} > R_\mathrm{c}^{(2)}$, where the number of total infected grows dramatically upon a small increase in the basic reproduction number $R_0$, quickly approaching the outbreak size expected without interventions. This transition is reflected in the strong variation of the time evolution (panel h) near $R_\mathrm{c}^{(3)}$. Dashed lines indicate the expected number of infected in the standard SIR model and with unlimited intervention capacity $\kappa \rightarrow \infty$, respectively. 
    All results are illustrated for total population size $N=10^6$, recovery rate $\gamma = 1$, $\delta = 1$, and intervention capacity $\kappa = 10^4$. 
    }
    \label{fig:FIG1_observation_explosive}
\end{figure*}

In the standard SIR model, susceptible (S) individuals become infected (I) with rate $\beta\,S\,I/N$ and are removed or recovered (R) with a rate $\gamma\,I$ (Fig.~\ref{fig:FIG1_observation_explosive}a-e). Here, we denote both the states and the absolute number of individuals in that state with capital letters $S$, $I$ and $R$. The parameters $\beta$ and $\gamma$ describe the infection rate per contact and the recovery rate per individual, respectively. The basic reproduction number $R_0 = \beta / \gamma$ quantifies the expected secondary infections caused by a single infected in a fully susceptible population and characterizes the qualitative collective dynamics of the model. If $R_0 < R_\mathrm{c}^{(1)} = 1$, the number of infected individuals $I(t)$ on average exponentially decreases with time $t$. If $R_0 > R_\mathrm{c}^{(1)}$, it initially increases exponentially. As a result, in the limit of an infinitely large population $N \rightarrow \infty$, a macroscopic outbreak occurs and ultimately affects some positive fraction $i_\mathrm{tot} = \lim_{N \rightarrow \infty} I_\mathrm{tot}/N > 0$ of the total population, where $I_\mathrm{tot} = N - \lim_{t \rightarrow \infty} S(t)$ describes the total number of individuals ever infected. This relative total outbreak size $i_\mathrm{tot}$ serves as an order parameter, distinguishing the two regimes, and is implicitly given by \cite{House2012ModellingEpidemicsNetworks}
\begin{equation}
    i_\mathrm{tot} = 1-e^{-R_0\,i_\mathrm{tot}} \,
\end{equation}
with a solution $i_\mathrm{tot} > 0$ only above a critical reproduction rate, $R_0 > R_\mathrm{c}^{(1)} = 1$, compare Fig.~\ref{fig:FIG1_observation_explosive}e.

We modify the standard SIR model to include capacity-limited interventions by adding a single new state (Q) (e.g. quarantine or treatment), see Fig.~\ref{fig:FIG1_observation_explosive}f-j for an illustration of this SIRQ model. 
In addition to the standard state transitions, infected individuals are removed into a state $Q$ at an additional rate $\delta\,I$ but at most at a rate $\kappa$, denoting the intervention capacity in units of individuals per time. The microscopic dynamics of both models follow a stochastic process where all transitions occur as independent Poisson processes at their given rates. These dynamics determine the probability $\mathcal{P}$ of an outbreak when a single individual is initially infected (compare Fig.~\ref{fig:FIG1_observation_explosive}d and i). The macroscopic dynamics in the limit of infinitely large populations, $N \rightarrow \infty$ are described by the mean field rate equations 
\begin{eqnarray}
    \frac{\mathrm{d}s}{\mathrm{d}t} &=& -\beta\,s\,i \label{eq:rate_equations}\\
    \frac{\mathrm{d}i}{\mathrm{d}t} &=& \beta\,s\,i - \gamma\,i - \mathrm{min}\left[\delta\,i, \lim_{N \rightarrow \infty} \kappa/N \right] \nonumber\,,
\end{eqnarray}
where the lower case letters $s$, $i$ and $r$ denote the fraction of individuals in the corresponding state, e.g. $s = \lim_{N \rightarrow \infty} S/N$. These dynamics govern the relative total outbreak size $i_\mathrm{tot}$ if a macroscopic outbreak occurs. We numerically illustrate our arguments and calculations for parameters $\gamma = \delta = 1$ for clarity of presentation and vary the infection rate $\beta$ to set $R_0$.

Compared to the standard SIR model, the interventions shift the critical point because infected individuals are additionally removed into state $Q$. Macroscopic outbreaks only occur when $R_0 > R_\mathrm{c}^{(2)} > 1$ (compare Fig.~\ref{fig:FIG1_observation_explosive}e and j). Once the reproduction number even slightly crosses a second threshold $R_\mathrm{c}^{(3)} > R_\mathrm{c}^{(2)}$, the total number of infected surges dramatically (Fig.~\ref{fig:FIG1_observation_explosive}j). In contrast to the smooth changes with the reproduction number in the standard SIR model (Fig.~\ref{fig:FIG1_observation_explosive}b,c), such explosive transitions may pose major challenges for predictability and control of epidemic dynamics. Small changes such as stochastic number fluctuations in finite size systems or small deviations in the reproduction number may yield large qualitative changes in the epidemic dynamics (Fig.~\ref{fig:FIG1_observation_explosive}h).

Do these capacity-limited interventions create a discontinuous transition in the epidemic dynamics? 
As long as  $I(t) < I_\mathrm{c} = \kappa/\delta$, infected individuals recover at an effective rate $\gamma_\mathrm{eff}\,I = (\gamma + \delta)\,I = 2\,I$. The expected dynamics of the system are described by an effective reproduction number $R_\mathrm{eff} = \beta / \gamma_\mathrm{eff} = R_0 / 2$. Consequently, we expect the critical point above which macroscopic outbreaks occur to be shifted to $R_\mathrm{c}^{(2)} = 2$. If at any time there are more infected individuals, $I(t) > I_\mathrm{c}$, the effective recovery rate reduces to $\gamma_\mathrm{eff}\,I = \gamma\,I + \kappa < 2\,I$. To understand how this change affects the epidemic dynamics, we consider the early microscopic spreading dynamics. With a single initially infected individual, the number of currently infected changes by $+1$ or $-1$ with each infection or recovery event, respectively. The probability for each event is proportional to the rates of the respective state transitions. Even if the number of infected should decrease on average and we would expect the epidemic to die out, there is a non-zero probability to reach any number of currently infected $I(t) \leq N$. The early dynamics if $I(t)$ ever becomes larger than $I_\mathrm{c}$ is thus equivalent to the Gambler's Ruin threshold crossing problem, see \cite{supp} for a more detailed description.
In the following, we reveal three distinct phase transitions depending on the scaling of the intervention capacity with the population size, $\kappa \propto N^\alpha$. 

\begin{figure}[ht]
    \centering
    \includegraphics[width=0.45\textwidth]{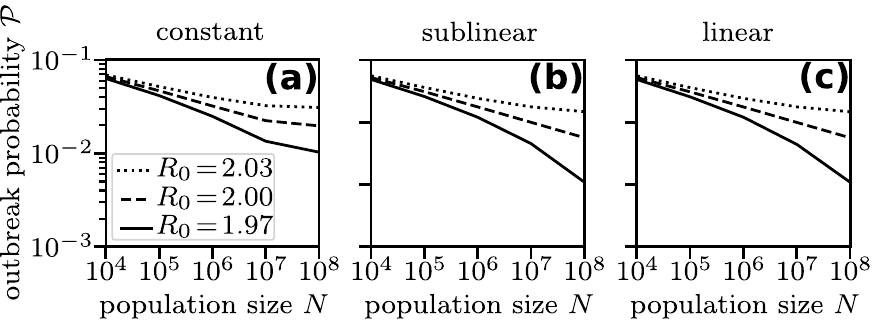}
    \caption{
    \textbf{Growing intervention capacity delays macroscopic outbreaks.} 
    The outbreak probability $\mathcal{P}$, computed as the fraction of realizations that reach $I_\mathrm{tot} > \sqrt{N}$ out of $10^5$ total realizations from one initially infected, $I_1(0) = 1$, behaves qualitatively differently for constant and (sub)linearly scaling intervention capacity. (a) For constant intervention capacity ($\kappa = 100$), the outbreak probability settles to non-zero values for any $R_0 > R_\mathrm{c}^{(1)} = 1$. (b,c) For sublinear and linear intervention capacity ($\kappa = \sqrt{N}$ and $\kappa = 0.01\,N$, respectively), the outbreak probability goes to zero for $R_0 < R_\mathrm{c}^{(2)} = 2$ in the limit of infinitely large populations; interventions prevent outbreaks. At the critical point $R_\mathrm{c}^{(2)} = 2$ the outbreak probability decreases as a power law as $N \rightarrow \infty$. 
    Note that due to defining outbreaks by $I_\mathrm{tot} > \sqrt{N}$, and our choice of $\kappa = \sqrt{N}$ and $\kappa = 0.01\,N \ge \sqrt{N}$ for $N \ge 10^4$, the outbreak probability for sublinear and linear scaling is identical since in both cases the intervention capacity is equal to or larger than our outbreak threshold.
    All results are shown for recovery rate $\gamma = 1$ and $\delta = 1$. 
\label{fig:FIG2_outbreak_probabilities}
}
\end{figure}

\begin{figure*}[ht]
    \centering
    \includegraphics[width=1.0\textwidth]{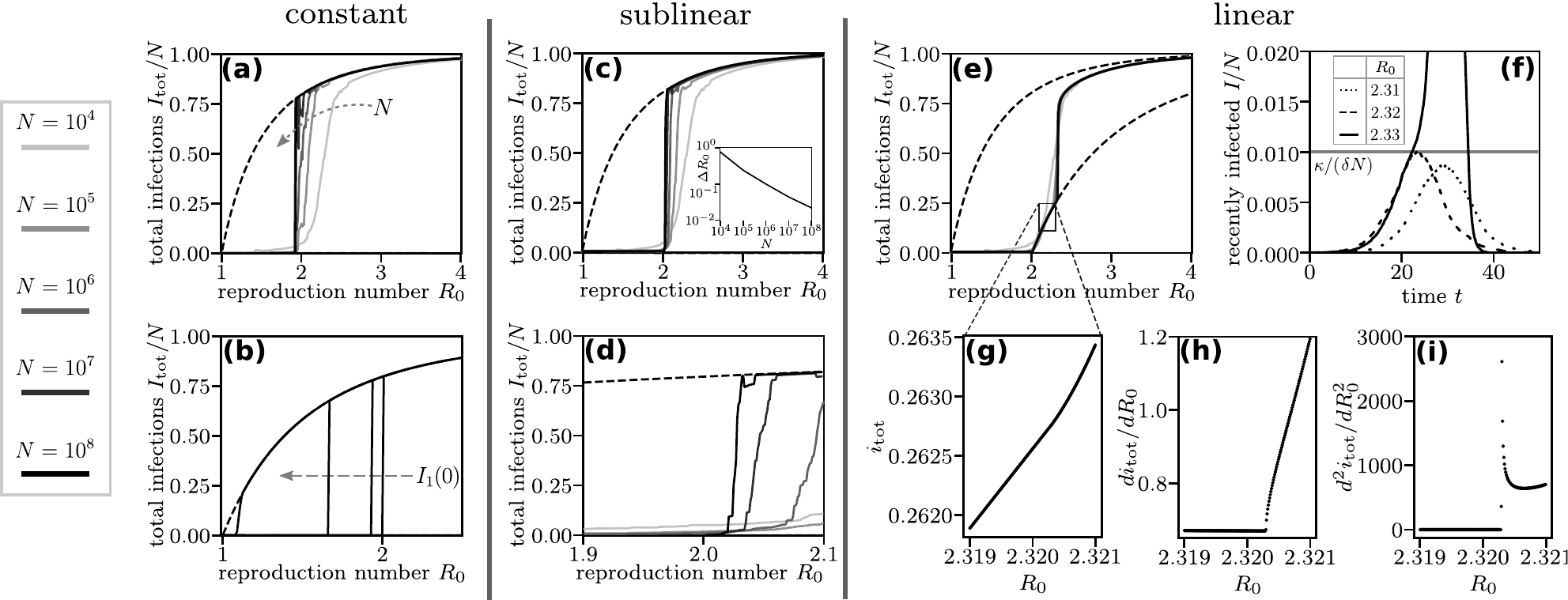}
    \caption{
        \textbf{Limited quarantine induces different explosive transitions.} 
        (a,b) For constant intervention capacity $\kappa = 100$, only few outbreaks are observed for $R_0 < R_\mathrm{c}^{(2)} = 2$ due to the small outbreak probability (panel a, compare Fig.~\ref{fig:FIG2_outbreak_probabilities}a). Outbreaks that do occur quickly grow to the same size as for standard SIR dynamics without interventions (dashed line), as also demonstrated by increasing the number of initially infected ($I_1(0) \in \{1, 10, 100, 1000\}$ for $N = 10^8$, panel b). 
        (c,d) For sublinear intervention capacity, $\kappa = N^{1/2}$, no outbreaks occur for $R_0 < R_\mathrm{c}^{(2)} = 2$ in the thermodynamic limit (compare Fig.~\ref{fig:FIG2_outbreak_probabilities}b). 
        For $R_0 > R_\mathrm{c}^{(2)}$, the outbreak size is close to that of SIR dynamics without intervention (dashed line). At the critical point $R_0 = R_\mathrm{c}^{(2)}$, the outbreak size increases discontinuously in the thermodynamic limit $N\rightarrow\infty$. (c, inset) The width $\Delta R_0 = R_0^{+} - R_0^{-}$ of the transition region in which the total number of infected increases from $I_\mathrm{tot}(R_0^{-})/N = 0.1$ to $I_\mathrm{tot}(R_0^{+})/N = 0.75$ decays to zero as $N \rightarrow \infty$.
        (e,f) For linear intervention capacity $\kappa = 0.01\,N$, no outbreaks occur for $R_0 < R_\mathrm{c}^{(2)} = 2$ in the thermodynamic limit (compare Fig.~\ref{fig:FIG2_outbreak_probabilities}c). 
        Above $R_\mathrm{c}^{(2)} = 2$ the outbreak size is initially the same as in the standard SIR model with increased effective recovery rate $\gamma + \delta$ (compare Fig.~\ref{fig:FIG1_observation_explosive}j). At a second critical point $R_\mathrm{c}^{(3)} \approx 2.3203$, where the concurrent number of infected during the outbreak overwhelms the intervention capacity ($I(t) > \kappa / \delta$, panel f), the outbreak size undergoes a second, sudden but continuous transition. 
        (g-i) The fraction of total infected $i_\mathrm{tot}$ computed from the mean-field rate equation Eq.~\eqref{eq:rate_equations} and its derivatives reveal a continuous third-order transition at $R_\mathrm{c}^{(3)}$ where only the second derivative $d^2i_\mathrm{tot}/dR_0^2$ is discontinuous. 
        All outbreak sizes are evaluated as averages over large outbreaks with $I_\mathrm{tot} > \sqrt{N}$ over at least $100$ realizations with $I_1(0) = 10$ initially infected (unless explicitly stated otherwise). All results are shown for recovery rate $\gamma = 1$ and $\delta = 1$.
        }
        \label{fig:FIG3_outbreak_sizes}
\end{figure*}

\paragraph{Constant intervention capacity.} 
For constant $\kappa \propto N^0$, the system exhibits some positive probability of reaching $I > I_\mathrm{c}$.  While this probability is exponentially suppressed with increasing intervention capacity $\kappa$, the interventions cannot completely prevent outbreaks. Once the number of infected becomes sufficiently large, the constant intervention rate $\kappa$ becomes negligible compared to the natural recovery rate $\gamma\,I$ and the system behaves like a standard SIR model without interventions. Consequently, macroscopic outbreaks occur with positive probability as soon as $R_0 > R_\mathrm{c}^{(1)} = 1$, similar to the standard SIR model, but the outbreak probability is exponentially suppressed with the intervention capacity $\kappa$ (Fig.~\ref{fig:FIG2_outbreak_probabilities}a). The macroscopic dynamics of the outbreaks that do occur is determined by the rate equations \eqref{eq:rate_equations}. The size of an outbreak (when it does occur) is thus the same as in the standard SIR model (compare Fig.~\ref{fig:FIG3_outbreak_sizes}a,b).

\paragraph{Sublinear intervention capacity.} 
For sublinearly scaling intervention capacity $\kappa \propto N^\alpha$ with $0 < \alpha < 1$, the same argument for the microscopic dynamics applies. However, now the threshold value $I_\mathrm{c} = \kappa / \delta \propto N^\alpha$ grows with the population size. Thus, the probability for the outbreak to grow beyond this threshold becomes zero in the large population limit $N \rightarrow \infty$ as long as the effective reproduction number $R_\mathrm{eff} = R_0 / 2 < 1$, i.e. as long as $R_0 < R_\mathrm{c}^{(2)} = 2$. Only then can macroscopic outbreaks occur with a finite probability (Fig.~\ref{fig:FIG2_outbreak_probabilities}b). When a macroscopic outbreak does occur, the intervention rate becomes negligible since it scales sublinearly with the population size, $\kappa/N \rightarrow 0$ as $N \rightarrow \infty$. The dynamics is equivalent to the standard SIR model. Consequently, we observe a discontinuous transition of the outbreak size at $R_\mathrm{c}^{(2)} = 2$ (Fig.~\ref{fig:FIG3_outbreak_sizes}c,d).

\paragraph{Linear intervention capacity} 
For linear intervention capacity $\kappa = \tilde{\kappa}\,N \propto N$ with constant $\tilde{\kappa}$, the dynamics become more intriguing. Again, the same argument for the microscopic dynamics applies as for the sublinearly scaling intervention capacity. Macroscopic outbreaks are only possible for $R_0 > R_\mathrm{c}^{(2)} = 2$ (Fig.~\ref{fig:FIG2_outbreak_probabilities}c). However, sufficiently small macroscopic outbreaks do not immediately exceed the intervention capacity threshold $I_\mathrm{c} \propto N$, and $I$ remains smaller than $I_\mathrm{c}$ during the outbreak. We thus observe a continuous second order transition equivalent to an SIR model with recovery rate $\gamma_\mathrm{eff} = (\gamma + \delta) = 2$. Only if the reproduction number is larger than a second critical value, $R_0 > R_\mathrm{c}^{(3)}$, the concurrently infected exceed the threshold $I_\mathrm{c}$ during the outbreak and a second transition occurs.

To reveal the type of the second transition, we compute the scaling of the number of additionally infected when the intervention capacity is overwhelmed. We here sketch the main steps in the argument, a step-by-step calculation is provided in the Supplemental Material \cite{supp}. We focus on the time $t^*$ at which the number of infected first exceeds the threshold $I(t^*) = I_\mathrm{c}$, or equivalently $i(t^*) = I_\mathrm{c}/N = \tilde{\kappa} / \left(N \delta\right)$. Until $t^*$, the dynamics are identical to a system with infinite intervention capacity $\kappa \rightarrow \infty$ with $i_\infty(t)$ infected. Exactly at $t^*$ both $i(t^*)$ and its first derivative are still the same as for infinite intervention capacity [Eq.~\eqref{eq:rate_equations}], but the second derivative changes to the right of $t^*$. Compared to a system with infinite intervention capacity we thus find
\begin{eqnarray}
    i(t) - i_\infty(t) &\sim \frac{1}{2}\left[i^{\prime\prime}(t^*) - i_\infty^{\prime\prime}(t^*)\right]\,\left(t-t^*\right)^2 \nonumber \\
    &\propto \left(R_0 - R_\mathrm{c}^{(3)}\right)^{1/2} \, \left(t-t^*\right)^2
\end{eqnarray}
However, the number of infected remains above the threshold only for a short time $\Delta t \propto \left(R_0 - R_\mathrm{c}^{(3)}\right)^{1/2}$, following from the quadratic expansion around the maximum of $i(t)$ which increases linearly with $\left(R_0 - R_\mathrm{c}^{(3)}\right)$ (see Supplemental Material \cite{supp} for details). We then find the leading order scaling of the additional infections by integrating the additional infection rate $\beta\,s(t) \,\left[i(t) - i_\infty(t)\right]$ for the time $\Delta t$, resulting in a leading order correction proportional to $\left[i^{\prime\prime}(t^*) - i_\infty^{\prime\prime}(t^*)\right]\,\Delta t^3 \propto \left(R_0 - R_\mathrm{c}^{(3)}\right)^2$. Secondary infections enter only as higher order corrections. We thus find that the second derivative of $i_\mathrm{tot}(R_0)$ is discontinuous at $R_0 = R_\mathrm{c}^{(3)}$ and the transition is sudden, yet third order, and thus surprisingly even smoother than the second order transition at $R_\mathrm{c}^{(2)}$ (Fig.~\ref{fig:FIG3_outbreak_sizes}e-i). 

The above explanations remain qualitatively valid for a broad class of systems since they only rely on scaling arguments to understand the impact of the limited intervention capacity on the microscopic dynamics and generic leading order behavior for the effect on the macroscopic dynamics. Our argument only requires that: (i) The system exhibits non-trivial outbreak dynamics even with infinite intervention capacity, ensuring that outbreaks exist in the first place if the intervention capacity scales with the population size. (ii) The intervention capacity enters the macroscopic dynamics as a hard limit such that the derivative of $i(t)$ is continuous but not differentiable when the number of infected overwhelms the intervention capacity [compare Eq.\eqref{eq:rate_equations}]. We provide a range of simple and more complex model variations illustrating these conditions and the robustness of the reported transitions in the Supplemental Material \cite{supp}.

Overall, our results offer a novel perspective on epidemic containment with capacity-limited countermeasures. The different types of explosive transitions to large outbreaks present different challenges for the predictability and control of epidemic dynamics. This applies in particular to the evaluation of containment measures across cities or countries when the intervention capacity depends on the population size. They also highlight the option of novel types of simultaneously explosive as well as third order phase transitions in complex systems in general. 

We acknowledge support from the Volkswagen Foundation under Grant.~No.~99~720 and the German Federal Ministry for Education and Research (BMBF) under grant number 16ICR01.
This research was supported by the Deutsche Forschungsgemeinschaft (DFG, German Research Foundation) under Germany´s Excellence Strategy – EXC-2068 – 390729961– Cluster of Excellence Physics of Life of TU Dresden. 

\nocite{zhonghua2010qualitative}

\bibliography{}

\end{document}


\global\long\def\pgr{\mathcal{P}_{\text{gr}}}
\global\long\def\pdb{\mathcal{P}_{\text{db}}}
\global\long\def\pov{\mathcal{P}_{\text{ov}}}
\global\long\def\pn{\mathcal{P}_{0}}
\global\long\def\df{d_{\text{f}}}

\title{
\setstretch{1.4}
Supplemental Material\\
{\small{accompanying the manuscript}} \\
Explosive Transitions in Epidemic Dynamics}

\author{Georg B\"orner}
\thanks{These authors contributed equally.}
\affiliation{
  Chair for Network Dynamics,
  Center for Advancing Electronics Dresden (cfaed) and Institute for Theoretical Physics, Technische Universität Dresden, 01062 Dresden, Germany
}

\author{Malte Schr\"oder}
\thanks{These authors contributed equally.}
\affiliation{
  Chair for Network Dynamics,
  Center for Advancing Electronics Dresden (cfaed) and Institute for Theoretical Physics, Technische Universität Dresden, 01062 Dresden, Germany
}

\author{Davide Scarselli}
\affiliation{Institute of Science and Technology Austria, Klosterneuburg, Austria}

\author{Nazmi Burak Budanur}
\affiliation{Institute of Science and Technology Austria, Klosterneuburg, Austria}
\affiliation{Max Planck Institute for the Physics of Complex Systems, Dresden, Germany}

\author{Björn Hof}
\affiliation{Institute of Science and Technology Austria, Klosterneuburg, Austria}

\author{Marc Timme}
\email{Correspondence should be addressed to: marc.timme@tu-dresden.de}
\affiliation{
  Chair for Network Dynamics,
  Center for Advancing Electronics Dresden (cfaed) and Institute for Theoretical Physics, Technische Universität Dresden, 01062 Dresden, Germany
}
\affiliation{
  Cluster of Excellence Physics of Life, Technical University of Dresden
  01062 Dresden, Germany
}
\affiliation{
  Lakeside Labs, Lakeside B04b, 9020 Klagenfurt, Austria
}

\maketitle
\onecolumngrid

\vspace{-0.8cm}

\section{Properties underlying the explosive transitions}

In the main manuscript we demonstrated diverse outbreak dynamics characterized by various qualitatively different explosive phase transitions as a function of the scaling of the intervention capacity with the population size. Here we illustrate the properties of the epidemic spreading models underlying our arguments and the observation in the main manuscript and how the dynamics change if these conditions do not hold.

While we illustrated the dynamics in the main manuscript for a single model (see section III of this Supplemental Material for a range of alternative model variations with qualitatively similar dynamics), our arguments explaining these dynamics only rely on two conditions that hold for a broad class of models:
\begin{itemize}
    \item non-trivial outbreak dynamics even with infinite intervention capacity
    \item an abrupt impact of the capacity limit in the macroscopic dynamics.
\end{itemize}
The first condition is required to observe any outbreaks at all when the intervention capacity increases with the population size. Otherwise, while outbreaks may grow if they are already macroscopic, no microscopic outbreak can grow sufficiently to become macroscopic in the first place. 

The second condition does not affect the dynamics with constant or sublinear intervention capacity since the interventions do not affect the macroscopic dynamics in these cases. However, it is central to the emergence of a secondary third-order transition when the intervention capacity scales linearly with the population size. 

\newpage

\subsubsection*{Trivial outbreak dynamics with interventions}

\begin{figure}[h]
    \centering
    \includegraphics{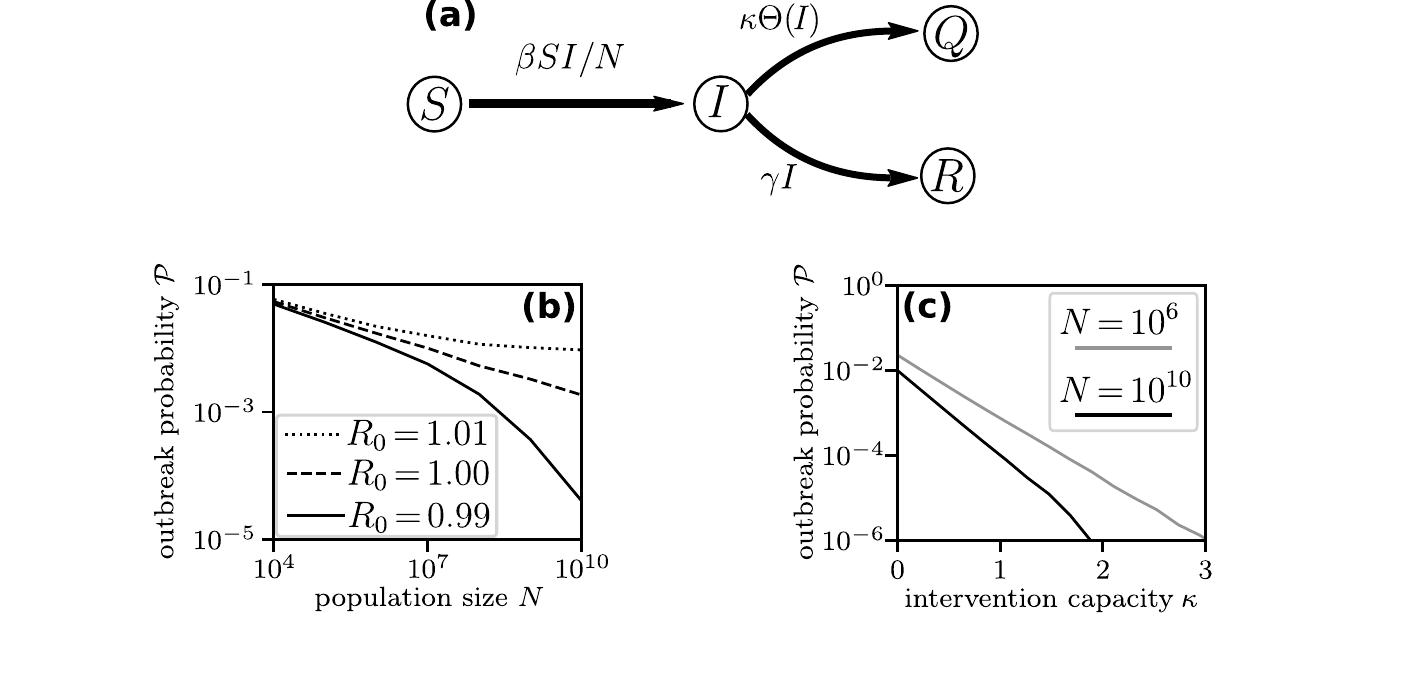}
    \caption{\textbf{Immediate targeted interventions prevent outbreaks}. (a) Compartment model sketch for the SIRQ model with instantaneous targeted intervention with total rate $\kappa$ independent of the number of infected individuals. $\Theta(\cdot)$ denotes the Heaviside step function.
    (b) Outbreak probability $P$ as a function of the population size $N$ for different basic reproduction numbers $R_0$ and constant intervention capacity $\kappa = 0.01$. For $R_0 < R_\mathrm{c}^{(1)} = 1$, $P$ decreases exponentially and outbreaks do not occur in the limit of an infinitely large population $N \to \infty$. For $R_0 > R_\mathrm{c}^{(1)} = 1$, $P$ takes on a small but constant value in the thermodynamic limit. 
    (c) Outbreak probability $P$ as a function of the intervention capacity $\kappa$ for fixed $R_0 = 1.01$. The outbreak probability is exponentially suppressed with increasing intervention capacity $\kappa$. Thus, if the intervention capacity increases with the population size, outbreaks are prevented for any $R_0$.
    The outbreak probability $\mathcal{P}$ is measured as the fraction of realizations that reach $I_\mathrm{tot} > \sqrt{N}$ out of $10^5$ (panel a) and $10^7$ (panel b) total realizations from one initially infected, $I_1(0) = 1$. As in the main manuscript, $\gamma = 1$.
         }
    \label{fig:SIRQ}
\end{figure}

First, we consider a model that violates the first condition and does not exhibit any outbreaks for infinite intervention capacity. We construct this model by implementing the intervention as immediate action with total rate $\kappa$, independent of the current number of infected (Fig.~\ref{fig:SIRQ}a). This may for example be interpreted as targeted interventions where individuals become infected, notice symptoms, and immediately get tested and treated or isolated without any delay. 

As the intervention capacity $\kappa$ grows, infected individuals are almost immediately removed and the number of infected cannot grow significantly as long as the inflow of new infected $\beta\,S\,I/N$ does not exceed the removal rate $\kappa$. For any constant (microscopic) number of infected $I = \mathcal{O}\left(1\right)$, in particular for one initial infected at $t=0$, this inflow is also microscopic $\beta\,S\,I/N = \mathcal{O}\left(1\right)$ since $S = \mathcal{O}\left(N\right)$. If $\kappa$ grows in any way with the total population size, in our context $\kappa \propto N^\alpha$ with $\alpha > 0$, this inflow can thus never exceed the intervention rate in the thermodynamic limit of infinitely large populations (and therefore $\kappa$ larger than any constant $\mathcal{O}\left(1\right)$). All infected are removed almost instantaneously, no new infections occur, and no (macroscopic) outbreak occurs, independent of the value of $R_0$. 

Outbreaks in the thermodynamic limit are thus only possible for constant intervention capacities $\kappa$ when (rare) random fluctuations may lead to sufficiently many infected to overwhelm the intervention. For $R_0 < 1$ and fixed intervention capacity $\kappa$, the outbreak probability $\mathcal{P}(R_0)$ becomes zero as $N \to \infty$, since macroscopic outbreaks never grow even without interventions. For $R_0 > 1$ it takes on a small but constant value, similar to the dynamics observed in the main manuscript (Fig.~\ref{fig:SIRQ}b). 

However, as a function of the intervention capacity $\kappa$, the outbreak probability is exponentially suppressed also for $R_0 > 1$ such that an increasing intervention capacity $\kappa \sim N^\alpha$ with $\alpha > 0$ prevents any outbreaks in the thermodynamic limit (Fig.~\ref{fig:SIRQ}c). 

These results illustrate why the first condition of nontrivial outbreak dynamics with infinite intervention capacity is required for our observations of the discontinuous transition and secondary third-order transition with sublinearly and linearly scaling intervention capacity, respectively. 

\newpage

\subsubsection*{Smooth impact of intervention capacity}
 
\begin{figure}[h]
    \centering
    \includegraphics{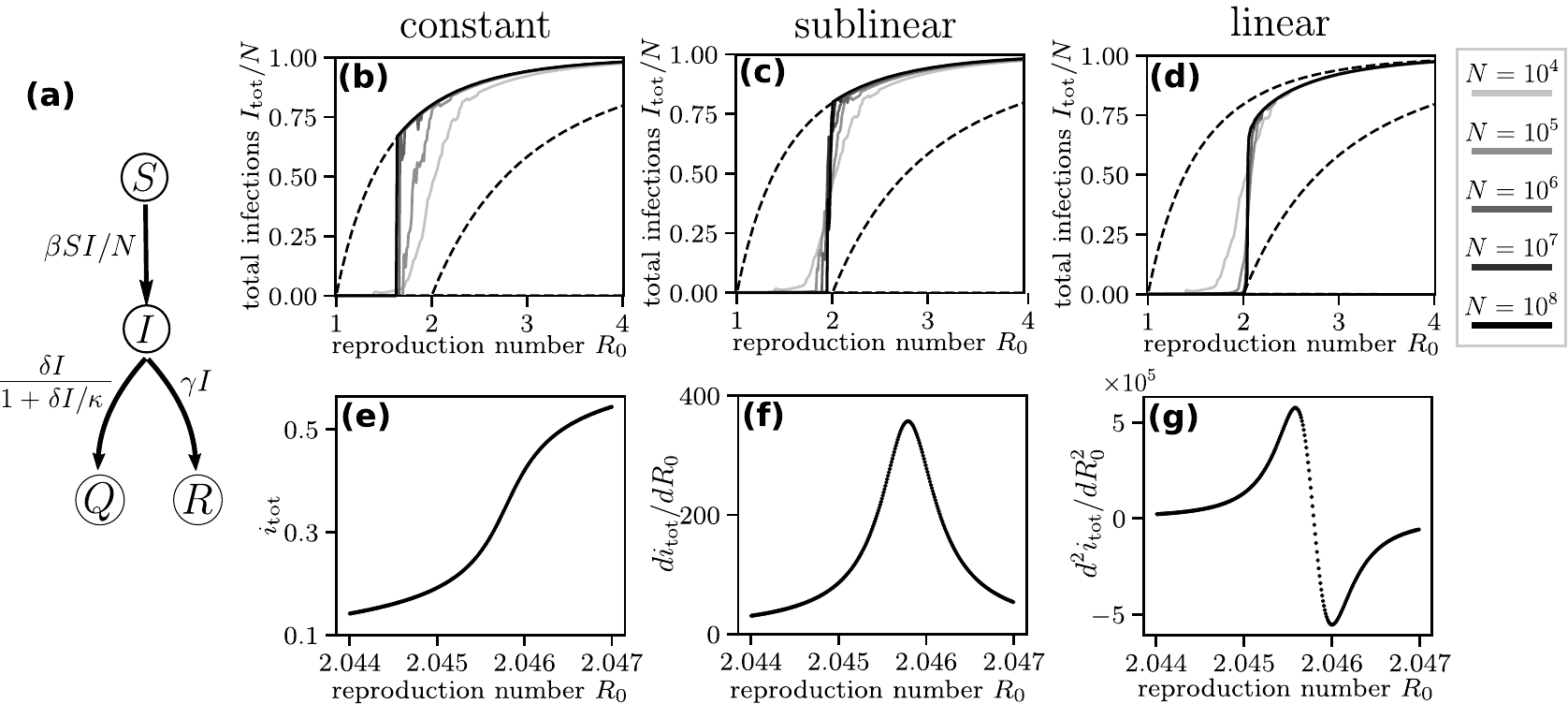}
        \caption{\textbf{Smooth impact of intervention capacity.} 
        (a) Compartment model sketch with smooth impact of the intervention capacity, Eq.~\eqref{eq:rate_equations_saturating_continuous}. 
        (b-d) Average outbreak size $I_\mathrm{tot}/{N}$ as a function of the basic reproduction number $R_0$ for intervention capacities $\kappa = 100$, $\kappa = \sqrt{N}$, and $\kappa = 0.01\,N$ (identical for $N = 10^4$ for all three scaling regimes). Dashed lines present standard SIR dynamics with effective recovery rates $\gamma = 1$ and $\gamma + \delta = 2$, respectively. We observe qualitatively the same outbreak dynamics for constant and sublinear scaling of the intervention capacity. 
        Outbreak sizes are averaged only over macroscopic outbreaks defined by $I_\mathrm{tot} > \sqrt{N}$ across a total of $100$ realizations with $I_1(0) = 10$ initially infected. 
        (e-g) Average outbreak size $i_\mathrm{tot}(R_0) = \lim_{N\rightarrow\infty} I_\mathrm{tot}(R_0) / N$ in the thermodynamic limit and its derivatives computed from mean-field differential equations Eq.~\eqref{eq:rate_equations_saturating_continuous} for linearly scaling intervention capacity $\kappa = 0.01\,N$ with a fraction of $i_1(0) = 10^{-6}$ initial infected. In contrast to the third-order transition observed in the main manuscript, the smooth saturation of the intervention rate as a function of the number of concurrently infected results in a continuous change of $i_\mathrm{tot}$ and all derivatives. However, the number of total infected still increases very suddenly (compare panel d and Fig.~3e in the main manuscript). All results are shown for $\gamma = 1$ and $\delta = 1$.
        }
    \label{fig:SIRQ_sat_cont}
\end{figure}

Second, we consider a model that violates the second condition and exhibits a smooth capacity limit in the macroscopic dynamics. Note that this is only relevant for linearly scaling intervention capacity $\kappa = \tilde{\kappa}\,N$ as otherwise the intervention capacity does not affect the macroscopic dynamics at all. We construct this model by replacing the hard limit in the time evolution equations, $\mathrm{min}\left[\delta i,\tilde{\kappa}\right]$, by a smooth function $\frac{\delta i}{1 + \delta i/\tilde{\kappa}}$ (Fig.~\ref{fig:SIRQ_sat_cont}a) such that
\begin{eqnarray}
    \frac{\mathrm{d}s}{\mathrm{d}t} &=& -\beta\,s\,i \label{eq:rate_equations_saturating_continuous}\\
    \frac{\mathrm{d}i}{\mathrm{d}t} &=& \beta\,s\,i - \gamma\,i - \frac{\delta\,i}{1 + \delta\,i/\tilde{\kappa}} \nonumber \,.
\end{eqnarray}
This model may be better interpreted in terms of testing capacity rather than intervention capacity after (unlimited) confirmed positive tests. For example, most tests and contact tracing attempts are negative when there are only few infections, $\frac{\delta\,i}{1 + \delta\,i/\tilde{\kappa}} \sim \delta\,i$ for $i \rightarrow 0$, similar to random testing in the model studied in the main manuscript. However, in contrast to the model in the main manuscript, the fraction of positive tests saturates smoothly as the number of infections grow, $\frac{\delta\,i}{1 + \delta\,i/\tilde{\kappa}} \sim \tilde{\kappa}$ as $\delta\,i/\tilde{\kappa} \rightarrow \infty$. Similar dynamics have been studied in \cite{zhonghua2010qualitative}.

Since the dynamics for a small number of infected in this model are identical to the SIRQ model studied in the main manuscript, we observe the same dynamics for constant intervention capacity and the same shift in the critical point $R_\mathrm{c}^{(2)}$ for (sub)linearly growing intervention capacity (Fig.~\ref{fig:SIRQ_sat_cont}b-d).

However, since the dynamics change smoothly as the concurrent number of infected increases, so does the total number of infected during an outbreak. In contrast to the discontinuous third derivative (compare Fig. 3g-i in the main manuscript), all derivatives of $i_\mathrm{tot} = \lim_{N \rightarrow \infty} I_\mathrm{tot}/N$ are continuous (Fig.~\ref{fig:SIRQ_sat_cont}e-g). Note that the transition is still explosive in the sense that a small change in the reproduction number causes a comparatively large change in the number of infected $I_\mathrm{tot}$ similar to the third order transition observed in the main manuscript. In single realizations of outbreaks in finite size systems (e.g. empirical observations), these different transitions may not be clearly distinguishable.

These results illustrate why the second condition of a hard impact of the intervention capacity is required for our observations of the secondary third-order transition with linearly scaling intervention capacity. We provide a more detailed analytical derivation of the third-order transition in section II of this Supplemental Material.

\clearpage

\section{Detailed explanation of the explosive transitions}
In the following, we discuss different analytical arguments supporting our observations in the main manuscript. In particular, we explain in more detail (i) the qualitative change in outbreak probability for constant intervention capacity and (ii) the location of the critical point $R_\mathrm{c}^{(2)}$ for (sub)linearly scaling intervention capacity by focussing on the microscopic stochastic dynamics with few infected individuals, as well as (iii) the emergence of the third-order transition for linearly scaling intervention capacity.

\subsection*{Microscopic dynamics and outbreak probability}
The probability of a macroscopic outbreak is determined by whether the epidemic dies out when there are only a small number of infected or whether the outbreak grows to infect a macroscopic fraction of the population. In the following, we consider the basic model discussed in the main manuscript. The same qualitative arguments also hold across model variations.

For a small number of concurrently infected $I < I_\mathrm{c} = \kappa / \delta$, the dynamics follows an SIR process with effective recovery rate $\gamma_\mathrm{eff} = \gamma + \delta$. Since we are interested in the initial stage of the outbreak, we assume that almost the whole population is still susceptible, $S \sim N$. The number of concurrently infected $I(t)$ then changes following a biased random walk with constant transition probabilities given by the relative probabilities of infection or recovery events, respectively,
\begin{eqnarray}
    I \, \rightarrow I + 1 \;\;:\;&\quad p_+& = \frac{\beta\,S\,I/N}{\beta\,S\,I/N + \gamma_\mathrm{eff}\,I} = \frac{R_0}{R_0 + \frac{\gamma + \delta}{\gamma}} \label{eq:supp_biased_random_walk_small}\\ 
    I \, \rightarrow I - 1 \;\;:\; &\quad p_-& = 1 - p_+ =  \frac{\gamma_\mathrm{eff}\,I}{\beta\,S\,I/N + \gamma_\mathrm{eff}\,I} = \frac{\frac{\gamma + \delta}{\gamma}}{R_0 + \frac{\gamma + \delta}{\gamma}} \nonumber\,.
\end{eqnarray}
with $R_0 = \beta / \gamma$ denoting the basic reproduction number without interventions. This description holds until the number of infected overwhelms the intervention capacity, $I > \kappa/\delta$. Above this threshold, the dynamics change and the effective recovery rate becomes $\gamma_\mathrm{eff}^> = \gamma + \kappa/I$, such that the dynamics are described by 
\begin{eqnarray}
    I \, \rightarrow I + 1 \;\;:\;&\quad p_+^>& = \frac{\beta\,S\,I/N}{\beta\,S\,I/N + \gamma_\mathrm{eff}^>\,I} = \frac{R_0}{R_0 + \frac{\gamma + \kappa/I}{\gamma}} > p_+ \label{eq:supp_biased_random_walk_large}\\ 
    I \, \rightarrow I - 1 \;\;:\; &\quad p_-^>& = 1 - p_+ =  \frac{\gamma_\mathrm{eff}^>\,I}{\beta\,S\,I/N + \gamma_\mathrm{eff}^>\,I} = \frac{\frac{\gamma + \kappa/I}{\gamma}}{R_0 + \frac{\gamma + \kappa/I}{\gamma}} < p_- \nonumber\,.
\end{eqnarray}
The random walk dynamics become self-reinforcing and more and more biased towards a growing number of infected as the number of infected increases, $\gamma_\mathrm{eff}^> \sim \gamma$ as $I \rightarrow \infty$.\\

For a single initial infected to cause sufficiently many infections to overwhelm the intervention capacity, the number of infected has to grow at least beyond $I_\mathrm{c} = \kappa / \delta$. The probability for this is given by the threshold crossing probability $p_{I_\mathrm{c}}(I)$ of the biased random walk Eq.~\eqref{eq:supp_biased_random_walk_small} when starting with $I$ infected. These threshold crossing probabilities are related by the self-consistency relation
\begin{eqnarray}
    p_{I_\mathrm{c}}(0) &=& 0 \\
    p_{I_\mathrm{c}}(1) &=& \left(1-p_+\right)\,p_{I_\mathrm{c}}(0) + p_+\,p_{I_\mathrm{c}}(2) \nonumber \\
    p_{I_\mathrm{c}}(2) &=& \left(1-p_+\right)\,p_{I_\mathrm{c}}(1) + p_+\,p_{I_\mathrm{c}}(3) \nonumber \\
    \vdots \nonumber 
\end{eqnarray}
equivalent to the well-known `Gambler's ruin'-problem. They resolve to 
\begin{equation}
    p_{I_\mathrm{c}}(1) = \frac{\frac{1-p_+}{p_+} - 1}{\left(\frac{1-p_+}{p_+}\right)^{I_\mathrm{c}} - 1} = \frac{\frac{\gamma + \delta}{\gamma\,R_0} -1}{\left(\frac{\gamma + \delta}{\gamma\,R_0} \right)^{\kappa/\delta} - 1} \,.
\end{equation}
In particular, this probability is always non-zero for any $R_0$ and constant intervention capacity $\kappa$. However, it decreases exponentially with $\kappa$ as long as $R_0 < \frac{\gamma + \delta}{\gamma}$. For large numbers of infected, the intervention capacity becomes negligible, and the outbreak dynamics become more similar to standard SIR dynamics described by  Eq.~\eqref{eq:supp_biased_random_walk_large} in the limit $I \rightarrow \infty$. An outbreak can thus grow into a macroscopic outbreak as long as $R_0 > 1$. 

\begin{figure}[h]
    \centering
     \includegraphics{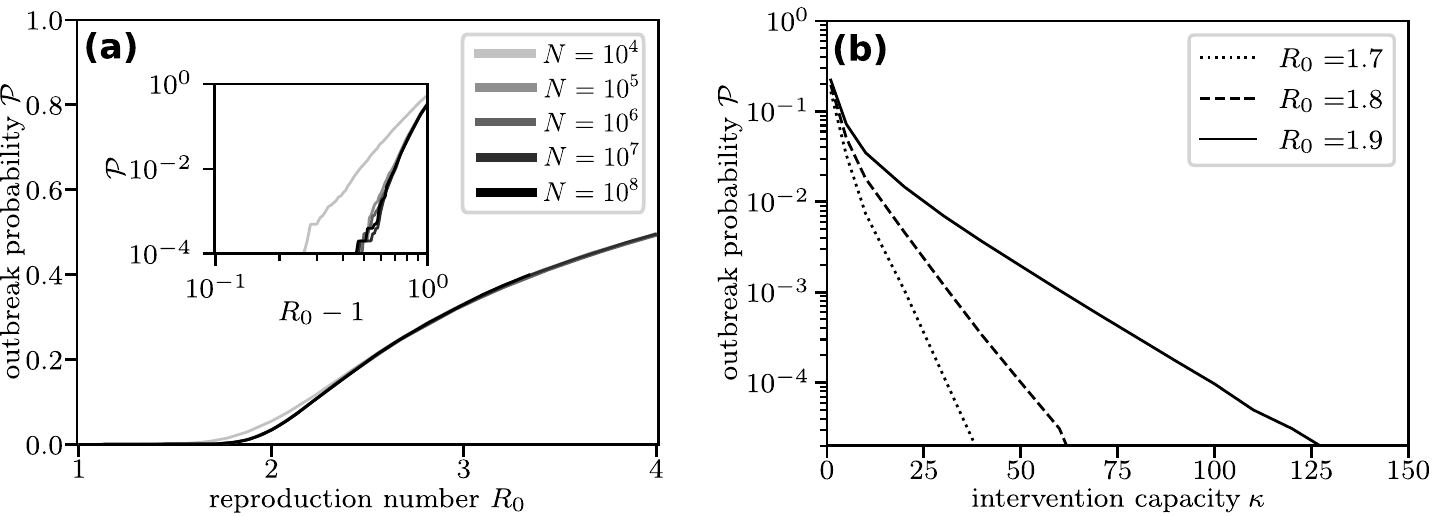}
    \caption{
        \textbf{Finite outbreak probability with constant intervention capacity.}
        (a) The outbreak probability with constant intervention capacity $\kappa = 20$ is small but positive for $R_0 < R_\mathrm{c}^{(2)} = 2$ and does not go to zero with increasing population size $N$ (compare inset). 
        (b) The probability of an outbreak decreases exponentially with the intervention capacity $\kappa$ but never reaches zero for finite $\kappa$ and $R_0 > R_\mathrm{c}^{(1)} = 1$, as illustrated for $N = 10^8$.
        The outbreak probability is measured as the fraction of realizations that reach $I_\mathrm{tot} > \sqrt{N}$ out of $10^4$ (panel a) and $10^6$ (panel b) total realizations from one initially infected, $I_1(0) = 1$. 
    }
    \label{fig:FIGS1_constant_quarantine_probabilty}
\end{figure}

Consequently, macroscopic outbreaks for constant intervention capacity occur as soon as $R_0 > R_\mathrm{c}^{(1)} = 1$, similar to the standard SIR model without interventions (Fig.~\ref{fig:FIGS1_constant_quarantine_probabilty}a). However, the probability for these outbreaks is exponentially suppressed as a function of the intervention capacity $\kappa$ (Fig.~\ref{fig:FIGS1_constant_quarantine_probabilty}b) as long as the random walk dynamics with a small number of infected are biased towards recovery events, i.e. as long as $p_+ < p_-$ or equivalently $R_0 < R_\mathrm{c}^{(2)} = \frac{\gamma + \delta}{\gamma}$. Beyond $R_0 > R_\mathrm{c}^{(2)}$, outbreaks grow with a non-zero probability regardless of the intervention.\\

Following the same arguments, when the intervention capacity grows with the population size, $\kappa \propto N^\alpha$ with $\alpha > 0$, macroscopic outbreaks are only possible for $R_0 > R_\mathrm{c}^{(2)} = \frac{\gamma + \delta}{\gamma}$. The probability to cross the threshold to overwhelm the intervention capacity decreases exponentially as the capacity grows and thus becomes zero in the thermodynamic limit of infinitely large populations. A macroscopic outbreak can thus only occur when already the dynamics with few infected Eq.~\eqref{eq:supp_biased_random_walk_small} are biased toward growing outbreaks, $R_0 > R_\mathrm{c}^{(2)} = \frac{\gamma + \delta}{\gamma}$. 

When an outbreak becomes macroscopic, the dynamics either (i) follow the standard SIR model when the intervention capacity scales sublinearly with the population size, $\kappa/I \rightarrow 0$ for $I = \mathcal{O}\left(N\right)$, or (ii) when the intervention capacity scales linearly with the population size, they follow the SIR dynamics with an effective recovery rate $\gamma + \delta$ until the number of concurrently infected overwhelms the intervention capacity during the outbreak and the secondary transition occurs discussed in more detail in the next section.\\

\newpage

\subsection*{Third-order phase transition of outbreak size}
When the intervention capacity scales linearly with the population size, $\kappa =\tilde{\kappa} \, N$, we observe a secondary third-order transition. We now study this transition in more detail and explain how it emerges, considering the basic model discussed in the main manuscript. The same qualitative arguments also hold across model variations. 

Since we are interested in the changes in the macroscopic dynamics, we focus on the rate equation description of the dynamics
\begin{eqnarray}
    \frac{\mathrm{d}s}{\mathrm{d}t} &=& -\beta\,i\,s \nonumber\\
    \frac{\mathrm{d}i}{\mathrm{d}t} &=& \beta\,i\,s - \gamma\,i - \mathrm{min}\left[\delta\,i, \tilde{\kappa}\right] \label{eq:supp_SIRQ_rate_equations}\\
    \frac{\mathrm{d}r}{\mathrm{d}t} &=& \gamma\,i + \mathrm{min}\left[\delta\,i, \tilde{\kappa}\right]\nonumber \,,
\end{eqnarray}
where the lower-case variables denote the relative number of agents in the corresponding state in the limit of an infinitely large total population, e.g. $s = \lim_{N \rightarrow \infty} S/N$. We are interested in the change of the total size of an outbreak, i.e. the fraction of individuals that was ever infected,
\begin{equation}
    i_\mathrm{tot} = \lim_{t \rightarrow \infty} r(t) = \lim_{t \rightarrow \infty}\,\left[1 - s(t) \right] 
\end{equation}
as a function of the basic reproduction number $R_0 = \beta/\gamma$. For sufficiently small $R_0 < R_\mathrm{c}^{(3)}$, the number of infected remains small and $\delta\,i < \tilde{\kappa}$ at all times such that the capacity limit of the interventions does not affect the dynamics. The evolution of the outbreak is then identical to the dynamics with infinite capacity $\tilde{\kappa} \rightarrow \infty$,
\begin{eqnarray}
    \frac{\mathrm{d}s_{\infty}}{\mathrm{d}t} &=& -\beta\,i_{\infty}\,s_{\infty} \nonumber\\
    \frac{\mathrm{d}i_{\infty}}{\mathrm{d}t} &=& \beta\,i_{\infty}\,s_{\infty} - (\gamma + \delta)\,i_{\infty} \label{eq:infinite_capacity}\\
    \frac{\mathrm{d}r_{\infty}}{\mathrm{d}t} &=& (\gamma + \delta)\,i_{\infty} \nonumber \,.
\end{eqnarray}
with the corresponding fraction of individuals ever infected,
\begin{equation}
    i^{\infty}_\mathrm{tot} = \lim_{t \rightarrow \infty} r_{\infty}(t) = \lim_{t \rightarrow \infty}\,\left[1 - s_{\infty}(t)\right] \,.
\end{equation}
These dynamics are exactly equivalent to a standard SIR model with effective recovery rate $\gamma_\mathrm{eff} = \gamma + \delta$, i.e. effective reproduction number $R_{\mathrm{eff}} = \frac{\beta}{\gamma_\mathrm{eff}} = R_0\,\frac{\gamma}{\gamma + \delta}$, explaining the delayed onset of outbreaks for $R_0 > R_\mathrm{c}^{(2)} = \frac{\gamma + \delta}{\gamma}$ (see also the discussion in the previous section).

\begin{figure}[h]
    \centering
    \includegraphics{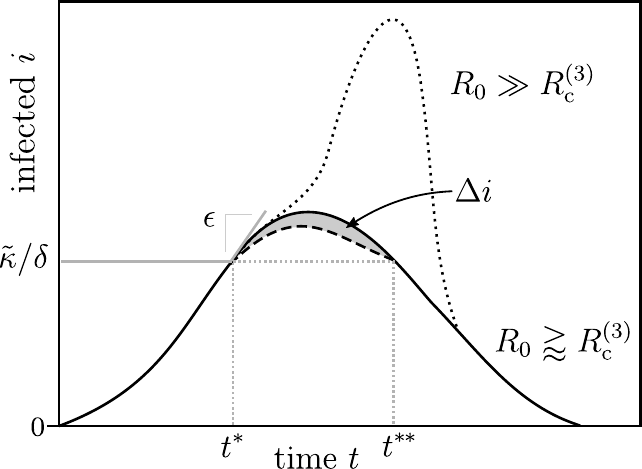}
    \caption{
        \textbf{Sketch of the dynamics just above the secondary transition $R_0 > R_\mathrm{c}^{(3)}$}. 
        Above the critical reproduction number, $R_0 \gtrapprox R_\mathrm{c}^{(3)}$, the fraction of infected $i(t)$ (solid line) crosses the threshold $\tilde{\kappa}/\delta$ at some time $t^*$. Up to $t^*$, the dynamics are identical to the system with infinite intervention capacity (dashed line), $i(t) = i_\infty(t)$, such that the slope at the threshold is also identical, $i^\prime(t^*) = i^\prime_{\infty}(t^*) = \epsilon > 0$. Beyond $t^*$, some infected are not removed by the intervention due to the limited capacity and cause additional secondary infections, $\Delta i(t) = i(t) - i_\infty(t) > 0$. The additional secondary infections are to leading order described by the integral $\int_{t^*}^{t^{**}} \beta s(t) \Delta i(t) \, dt > 0$ with $t^{**}$ denoting the time where the fraction of infected decreases below the threshold again (see main text and calculations below). 
        We study this transition by expanding the dynamics around the critical point $R_\mathrm{c}^{(3)}$ where the fraction of infected reaches but does not exceed the threshold, $\epsilon(R_\mathrm{c}^{(3)}) = 0$, $t^* = t^{**}$. 
        }
    \label{fig:FIGS2_linear_quarantine_expansion_sketch}
\end{figure}

The secondary transition occurs only for larger $R_0 = R_\mathrm{c}^{(3)}$ when $i(t)$ overwhelms the intervention capacity $\tilde{\kappa}$ during the outbreak. The total outbreak size increases compared to $i_\mathrm{tot}^{\infty}$ due to additional secondary infections by infected individuals that are not removed from $i$ due to the limited intervention capacity. The qualitative idea is sketched in Fig.~\ref{fig:FIGS2_linear_quarantine_expansion_sketch}. 

To understand how $i_\mathrm{tot}$ changes as $R_0$ increases beyond $R_\mathrm{c}^{(3)}$, we compute a leading order approximation
\begin{equation}
    i_\mathrm{tot}(R_0) = i_\mathrm{tot}^{\infty}(R_0) + \Delta\,i_\mathrm{tot}(R_0) \label{eq:supp_i_tot_expansion}
\end{equation}
for $R_0 > R_\mathrm{c}^{(3)}$ with $\Delta\,i_\mathrm{tot}\left(R_\mathrm{c}^{(3)}\right) = 0$. Following the idea sketched in Fig.~\ref{fig:FIGS2_linear_quarantine_expansion_sketch}, we expand the dynamics of $i(t)$ to the right of the point $t^*$ where $i(t^*) = \tilde{\kappa}/\delta$ just overwhelms the intervention capacity to compute the difference $\Delta i(t) = i(t) - i_\infty(t)$ to the reference state $i_{\infty}(t)$ with infinite intervention capacity. From these remaining infected, we finally compute the additional infections caused by the limited intervention capacity. We denote the small but positive slope of $i(t)$ at this point as $i^\prime(t^*) = \epsilon$. 

\newpage

Starting from these definitions, we compute all values and derivatives at $t^*$ (specifically for $t \rightarrow t^*$ from above) by solving and taking further derivatives of Eq.~\eqref{eq:supp_SIRQ_rate_equations} and \eqref{eq:infinite_capacity}. We find
\begin{eqnarray}
    s(t^*) = s_{\infty}(t^*) &=& \left(\frac{\delta}{\beta\,\tilde{\kappa}}\right)\,\epsilon + \left(\frac{\gamma + \delta}{\beta}\right) \label{eq:supp_s}\\[3mm]
    i(t^*) = i_{\infty}(t^*) &=& \frac{\tilde{\kappa}}{\delta} \label{eq:supp_i}\\[3mm]
    s^\prime(t^*) =  s_{\infty}^{\prime}(t^*) &=& - \epsilon - \left( \tilde{\kappa} + \frac{\gamma\,\tilde{\kappa}}{\delta}\right) \label{eq:supp_s_prime}\\[3mm]
    i^\prime(t^*) = i_{\infty}^{\prime}(t^*) &=& \epsilon \label{eq:supp_i_prime}\\[3mm]
    i^{\prime\prime}(t^*) &=& \left(\frac{\delta}{\tilde{\kappa}}\right)\,\epsilon^2 + \left(\delta - \frac{\beta\,\tilde{\kappa}}{\delta}\right)\,\epsilon + \left( - \frac{\beta\,\gamma\,\tilde{\kappa}^2}{\delta^2} - \frac{\beta\,\tilde{\kappa}^2}{\delta}\right)\\[3mm]
    i_{\infty}^{\prime\prime}(t^*) &=& \left(\frac{\delta}{\tilde{\kappa}}\right)\,\epsilon^2 + \left( -\frac{\beta\,\tilde{\kappa}}{\delta}\right)\,\epsilon + \left( - \frac{\beta\,\gamma\,\tilde{\kappa}^2}{\delta^2} - \frac{\beta\,\tilde{\kappa}^2}{\delta}\right)
\end{eqnarray}
Since the second derivative of $s(t)$ and $s_{\infty}(t)$ only depends on the first derivatives, it is the same with and without the intervention limit,
\begin{eqnarray}
    \Delta s^{\prime\prime}(t^*) &=& \frac{\mathrm{d}}{\mathrm{d}t}\left.\left(-\beta\,i(t)\,s(t) \right)\right|_{t=t^*} - \frac{\mathrm{d}}{\mathrm{d}t}\left.\left(-\beta\,i_{\infty}(t)\,s_{\infty}(t) \right)\right|_{t=t^*} \nonumber\\
                                 &=& \left.\left(-\beta\,i^\prime(t)\,s(t) -\beta\,i(t)\,s^\prime(t) \right)\right|_{t=t^*} - \left.\left(-\beta\,i_{\infty}^{\prime}(t)\,s_{\infty}(t) -\beta\,i_{\infty}(t)\,s_{\infty}^{\prime}(t) \right)\right|_{t=t^*} \nonumber\\
                                 &=& 0
\end{eqnarray}

\newpage

We expand the difference $\Delta i(t)$ to the right of $t^*$ up to third order as
\begin{equation}
    \Delta i(t) = \Delta i(t^*) + \Delta i^{\prime}(t^*)\,\left(t-t^*\right) + \frac{\Delta i^{\prime\prime}(t^*)}{2}\,\left(t-t^*\right)^2 + \frac{\Delta i^{\prime\prime\prime}(t^*)}{6}\,\left(t-t^*\right)^3 + \mathcal{O}\left[\left(t-t^*\right)^4\right] \,.
\end{equation}
The first two terms of this expansion are zero since the dynamics of both $i(t)$ and the reference state $i_{\infty}(t)$ are identical up to $t^*$ [compare Eq.~\eqref{eq:supp_s}-\eqref{eq:supp_i_prime}]. We compute the second and third term by directly evaluating the derivatives
\begin{eqnarray}
    \Delta i^{\prime\prime}(t^*) &=& i^{\prime\prime}(t^*) - i_\infty^{\prime\prime}(t^*) \nonumber\\
                                 &=& \frac{\mathrm{d}}{\mathrm{d}t}\left.\left(\beta\,i(t)\,s(t) - \gamma\,i(t) - \tilde{\kappa}\right)\right|_{t=t^*} - \frac{\mathrm{d}}{\mathrm{d}t}\left.\left(\beta\,i_{\infty}(t)\,s_{\infty}(t) - (\gamma + \delta)\,i_{\infty}(t)\right)\right|_{t=t^*} \nonumber\\
                                 &=& \left.\left(\beta\,i^\prime(t^*)\,s(t) + \beta\,i(t)\,s^\prime(t) - \gamma\,i^\prime(t) \right)\right|_{t=t^*} - \left.\left(\beta\,i_{\infty}^{\prime}(t)\,s_{\infty}(t) + \beta\,i_{\infty}(t)\,s_{\infty}^{\prime}(t) - (\gamma + \delta)\,i_{\infty}^{\prime}(t)\right)\right|_{t=t^*} \nonumber\\
                                 &=& \delta\,i_{\infty}^{\prime}(t^*) \nonumber\\
                                 &=& \delta\,\epsilon 
\end{eqnarray}
again using that the values and derivatives at $t^*$ are identical for $i(t)$ and the reference state $i_{\infty}(t)$. For the third term, we find
\begin{eqnarray}
    \Delta i^{\prime\prime\prime}(t^*) &=& i^{\prime\prime\prime}(t^*) - i_\infty^{\prime\prime\prime}(t^*) \nonumber\\
                                 &=& \frac{\mathrm{d}^2}{\mathrm{d}t^2}\left.\left(\beta\,i(t)\,s(t) - \gamma\,i(t) - \hat{\tilde{\kappa}}\right)\right|_{t=t^*} - \frac{\mathrm{d}^2}{\mathrm{d}t^2}\left.\left(\beta\,i_{\infty}(t)\,s_{\infty}(t) - (\gamma + \delta)\,i_{\infty}(t)\right)\right|_{t=t^*} \nonumber\\
                                 &=& \left.\left(\beta\,i^{\prime\prime}(t)\,s(t) + 2\beta\,i^\prime(t)\,s^\prime(t) + \beta\,i(t)\,s^{\prime\prime}(t) - \gamma\,i^{\prime\prime}(t) \right)\right|_{t=t^*} \nonumber\\
                                 &&\quad - \left.\left(\beta\,i_\infty^{\prime\prime}(t)\,s_{\infty}(t) + 2\beta\,i_{\infty}^{\prime}(t)\,s_{\infty}^{\prime}(t) + \beta\,i_{\infty}(t)\,s_{\infty}^{\prime\prime}(t) - (\gamma + \delta)\,i_{\infty}^{\prime\prime}(t)\right)\right|_{t=t^*} \nonumber\\
                                 &=& \beta\,\Delta i^{\prime\prime}(t^*)\,s(t^*) + \beta\,i(t^*)\,\Delta s^{\prime\prime}(t^*) - \gamma\,\Delta i^{\prime\prime}(t^*) + \delta\,i_{\infty}^{\prime\prime}(t^*) \nonumber\\
                                 &=& \beta\,\delta\,\epsilon\,s(t^*) - \gamma\,\delta\,\epsilon + \delta\,i^{\infty\prime\prime}(t^*) \nonumber\\
                                 &=& \beta\,\delta\,\epsilon\,\left(\left(\frac{\delta}{\beta\,\tilde{\kappa}}\right)\,\epsilon + \left(\frac{\gamma + \delta}{\beta}\right)\right) - \gamma\,\delta\,\epsilon + \delta\,\left(\left(\frac{\delta}{\tilde{\kappa}}\right)\,\epsilon^2 + \left( -\frac{\beta\,\tilde{\kappa}}{\delta}\right)\,\epsilon + \left( - \frac{\beta\,\gamma\,\tilde{\kappa}^2}{\delta^2} - \frac{\beta\,\tilde{\kappa}^2}{\delta}\right)\right) \nonumber\\
                                 &=& \left( \frac{2\,\delta^2}{\tilde{\kappa}} \right)\,\epsilon^2 + \left( \delta^2 - \beta\tilde{\kappa} \right)\,\epsilon + \left(- \frac{\beta\,\gamma\,\tilde{\kappa}^2}{\delta} - \beta\,\tilde{\kappa}^2\right) \label{eq:supp_delta_i_3prime}
\end{eqnarray}
We thus find $\Delta i(t)$ up to third order in $\left(t - t^*\right)$  as
\begin{eqnarray}
    \Delta i(t) &=& \frac{\delta\,\epsilon}{2}\,\left(t-t^*\right)^2 + \frac{1}{6}\,\left[\left( \frac{2\,\delta^2}{\tilde{\kappa}} \right)\,\epsilon^2 + \left( \delta^2 - \beta\tilde{\kappa} \right)\,\epsilon + \left(- \frac{\beta\,\gamma\,\tilde{\kappa}^2}{\delta} - \beta\,\tilde{\kappa}^2\right)\right]\,\left(t-t^*\right)^3 + \mathcal{O}\left[\left(t-t^*\right)^4\right] \\
    &=& a_2 \, \left(t-t^*\right)^2 + a_3\,\left(t-t^*\right)^3 + \mathcal{O}\left[\left(t-t^*\right)^4\right] \nonumber
\end{eqnarray}
with $a_2 = \mathcal{O}\left(\epsilon\right)$ and $a_3 = \mathcal{O}\left(1\right)$.
In particular, we observe that $\Delta i^{\prime\prime\prime}(t^*) < 0$ for small $\epsilon$ such that $\Delta i(t)$ first increases quadratically but then quickly vanishes again.

\newpage

The total additional infections are mainly the result of the direct secondary infections caused by the infected $\Delta i(t)$, described by the following integral
\begin{equation}
    \Delta\,i_\mathrm{tot}(R_0) = \mathcal{O} \left[\, \int_{t^*}^{t^{**}} \beta\,\Delta i(t)\,s(t^*)\,\mathrm{d}t\, \right]
\end{equation}
until the time $t^{**}$ where $\Delta i(t)$ becomes zero again. We find the relevant time difference $\left(t^{**} - t^*\right)$ as 
\begin{eqnarray}
    0 &=& \frac{\delta\,\epsilon}{2}\,\left(t^{**}-t^*\right)^2 + \frac{1}{6}\,\left[\left( \frac{2\,\delta^2}{\tilde{\kappa}} \right)\,\epsilon^2 + \left( \delta^2 - \beta\tilde{\kappa} \right)\,\epsilon + \left(- \frac{\beta\,\gamma\,\tilde{\kappa}^2}{\delta} - \beta\,\tilde{\kappa}^2\right)\right]\,\left(t^{**}-t^*\right) \nonumber\\
    \left(t^{**}-t^*\right) &=& -\frac{3 \delta\,\epsilon}{\left( \frac{2\,\delta^2}{\tilde{\kappa}} \right)\,\epsilon^2 + \left( \delta^2 - \beta\tilde{\kappa} \right)\,\epsilon + \left(- \frac{\beta\,\gamma\,\tilde{\kappa}^2}{\delta} - \beta\,\tilde{\kappa}^2\right)} \label{eq:supp_integration_time_diff}\\
    &=& \mathcal{O}\left(\epsilon\right) \nonumber \, ,
\end{eqnarray}
scaling linearly in $\epsilon$. The integral therefore scales as $\epsilon^4$, by evaluating the relevant terms
\begin{eqnarray}
    \int_{t^*}^{t^{**}} \beta\,\Delta i(t)\,s(t^*)\,\mathrm{d}t \nonumber &\sim&  \int_{t^*}^{t^{**}} \beta\,\left[a_2 \, \left(t-t^*\right)^2 + a_3\,\left(t-t^*\right)^3\right]\,\left[\frac{\gamma + \delta}{\beta}\right]\,\mathrm{d}t \nonumber\\
    &\sim&  \frac{\left(\gamma + \delta\right)\,a_2}{3} \, \left(t^{**}-t^*\right)^3 + \frac{\left(\gamma + \delta\right)\,a_3}{4} \, \left(t^{**}-t^*\right)^4 \label{eq:supp_additional_infected_integral}\\
    &=& \mathcal{O}\left(\epsilon^4\right) \nonumber\,,
\end{eqnarray}
since $a_2 = \mathcal{O}\left(\epsilon\right)$, $a_3 = \mathcal{O}\left(1\right)$, and $\left(t^{**} - t^*\right) = \mathcal{O}\left(\epsilon\right)$. All other terms, such as the correction to $s(t^*)$ or accounting for the fact that $s(t)$ actually decreases, only contribute to higher orders in $\epsilon$.

To transfer this result to the proper scaling with $R_0 - R_\mathrm{c}^{(3)}$, we now determine the dependence of $\epsilon$ on $R_0 - R_\mathrm{c}^{(3)}$. During the outbreak, the number of infected $i(t)$ attains a maximum at some time $t_\mathrm{max}$. Expanding $i$ around this maximum at $t_\mathrm{max}$, we find
\begin{eqnarray}
    i(t) &=& i_\mathrm{max}\left(R_0 - R_\mathrm{c}^{(3)}\right) - \frac{c_2}{2}\,\left(t - t_\mathrm{max}\right)^2 + \mathcal{O}\left[\left(t - t_\mathrm{max}\right)^3\right] \nonumber\\
    &=& \frac{\kappa}{\delta} + c_1\,\times\,\left(R_0 - R_\mathrm{c}^{(3)}\right) - \frac{c_{2,0}}{2}\,\left(t - t_\mathrm{max}\right)^2 + \mathcal{O}\left[\left(R_0 - R_\mathrm{c}^{(3)}\right)^2\right] + \mathcal{O}\left[\left(t - t_\mathrm{max}\right)^3\right] \,,
\end{eqnarray}
where we expand $i_\mathrm{max}\left(R_0 - R_\mathrm{c}^{(3)}\right)$ and the second coefficient $c_2$ around $R_\mathrm{c}^{(3)}$ to linear order such that $c_1$ and $c_{2,0}$ do not depend on $R_0$ but only on $R_\mathrm{c}^{(3)}$. Evaluating this expression at $t^*$ where $i(t^*) = \frac{\tilde{\kappa}}{\delta}$, we find 
\begin{equation}
    \left(t^* - t_\mathrm{max}\right)^2 = \frac{2 c_0\left(R_0 - R_\mathrm{c}^{(3)}\right) }{c_{2,0}} \,.
\end{equation}
Now, solving the derivative $i^\prime(t^*) = \epsilon$ reveals that $\epsilon$ scales as
\begin{eqnarray}
    \epsilon = i^\prime(t^*) &=& -c_2\,\left(t^* - t_\mathrm{max}\right) +\mathcal{O}\left[\left(t^* - t_\mathrm{max}\right)^2\right] \nonumber\\
    &=& \sqrt{2\,c_0\,c_{2,0}\,\left(R_0 - R_\mathrm{c}^{(3)}\right)} + \mathcal{O}\left[R_0 - R_\mathrm{c}^{(3)}\right] \\
    &=& \mathcal{O} \left[ \left(R_0 - R_\mathrm{c}^{(3)}\right)^{1/2} \right] \nonumber\,.
\end{eqnarray}

Combining these results, we thus find that the additional number of infected scales to leading order quadratically in the difference $\left(R_0 - R_\mathrm{c}^{(3)}\right)$,
\begin{equation}
    \Delta\,i_\mathrm{tot}(R_0) = \mathcal{O}\left[\left(R_0 - R_\mathrm{c}^{(3)}\right)^{2}\right]   \quad\quad\mathrm{for}\quad R_0 > R_\mathrm{c}^{(3)}.
\end{equation}
Overall, the total number of infected thus behaves like
\begin{eqnarray}
    i_\mathrm{tot}(R_0) &=& i_\mathrm{tot}^{\infty}(R_0) + \Delta\,i_\mathrm{tot}(R_0) \nonumber\\
    &=& i^{\infty}_\mathrm{tot}(R_0) + K\,\left(R_0 - R_\mathrm{c}^{(3)}\right)^2 + \mathcal{O}\left[\left(R_0 - R_\mathrm{c}^{(3)}\right)^{2.5}\right],
\end{eqnarray}
with some constant $K > 0$ for $R_0 > R_\mathrm{c}^{(3)}$. The secondary transition is thus a third-order transition where the observable and its first derivative are continuous, but the second derivative is discontinuous at $R_\mathrm{c}^{(3)}$. This smoothness of the transition stands in stark contrast to the initial observation of a very sudden, explosive transition that is explained by the prefactor $K$. For example, the integration time $t^{**} - t^*$ in Eq.~\eqref{eq:supp_integration_time_diff} scales as $\kappa^{-2}$, thus entering as $\kappa^{-6}$ in the additional infected Eq.~\eqref{eq:supp_additional_infected_integral}. The prefactor thus strongly depends on the intervention capacity, explaining the sudden onset of the transition especially for small $\tilde{\kappa}$.\\

All calculations above cover the leading order of the correction due to the limited intervention capacity and rely only on the generic scaling of the quantities involved and their expansion around the point where the intervention capacity is overwhelmed for the first time. Importantly, the same (qualitative) calculation describes the dynamics of a broad class of epidemic models with limited intervention capacity $\kappa = \tilde{\kappa}\,N$. The only conditions are, as described above: (i) The system exhibits non-trivial outbreak dynamics even with infinite intervention capacity, such that macroscopic outbreaks occur for sufficiently large $R_0 > R_\mathrm{c}^{(2)}$, in order to observe any outbreaks at all when the intervention capacity scales with the population size. (ii) The intervention capacity enters the macroscopic dynamics as a hard limit such that the derivative of $i(t)$ is continuous but not differentiable at the point where the number of infected overwhelm the intervention capacity. The second condition is particularly relevant as the change in the second derivative of $i(t)$ at $t^*$ compared to $i^\infty(t)$ directly results in the leading order of the correction computed above. In contrast, if the dynamics change smoothly as $i(t)$ increases, the total infected $i_\mathrm{tot}$ also change smoothly with the reproduction number $R_0$ (compare section I of this Supplemental Material).\\

\begin{figure}[h]
    \centering
     \includegraphics{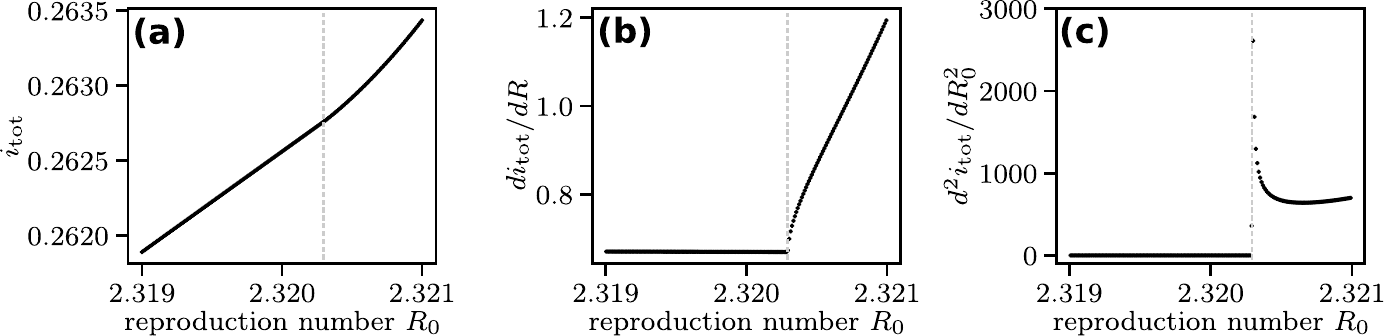}
    \caption{
        \textbf{Third-order transition with linear intervention capacity.}
        (a,b) The fraction of total infected $i_\mathrm{tot}$ and its first derivative with respect to $R_0$ are both continuous at the secondary transition point $R_\mathrm{c}^{(3)}$ (vertical dashed line). (c) The second derivative is discontinuous, illustrating the third-order transition. Results are shown for $\gamma = 1$, $\delta = 1$, and $\kappa = 0.01\,N$ as in the main manuscript. 
    }
    \label{fig:FIGS3_linear_quarantine_expansion_numerics}
\end{figure}

Fig.~\ref{fig:FIGS3_linear_quarantine_expansion_numerics} illustrates the total fraction of infected $i_\mathrm{tot}$ from the numerical solutions of the rate equations in the vicinity of the second transition, $R_0 \approx R_\mathrm{c}^{(3)}$ (compare Fig.~3g-i in the main manuscript). Both $i_\mathrm{tot}$ and its first derivative with respect to $R_0$ are continuous, the second derivative is discontinuous, illustrating the third-order transition derived above.

\clearpage

\section{Robustness of explosive transitions across model variations}

In the following, we consider various modifications and extensions of the SIRQ model analyzed in the main manuscript. All model variations analyzed here fulfill the conditions described in the first part of this Supplemental Material and represent different intervention dynamics. These models demonstrate that the observations made in the main manuscript hold for a wide class of systems, including also more realistic disease progression or intervention dynamics.

 \subsection*{Delayed intervention and varied disease progression}

\begin{figure}[h]
    \centering
    \includegraphics{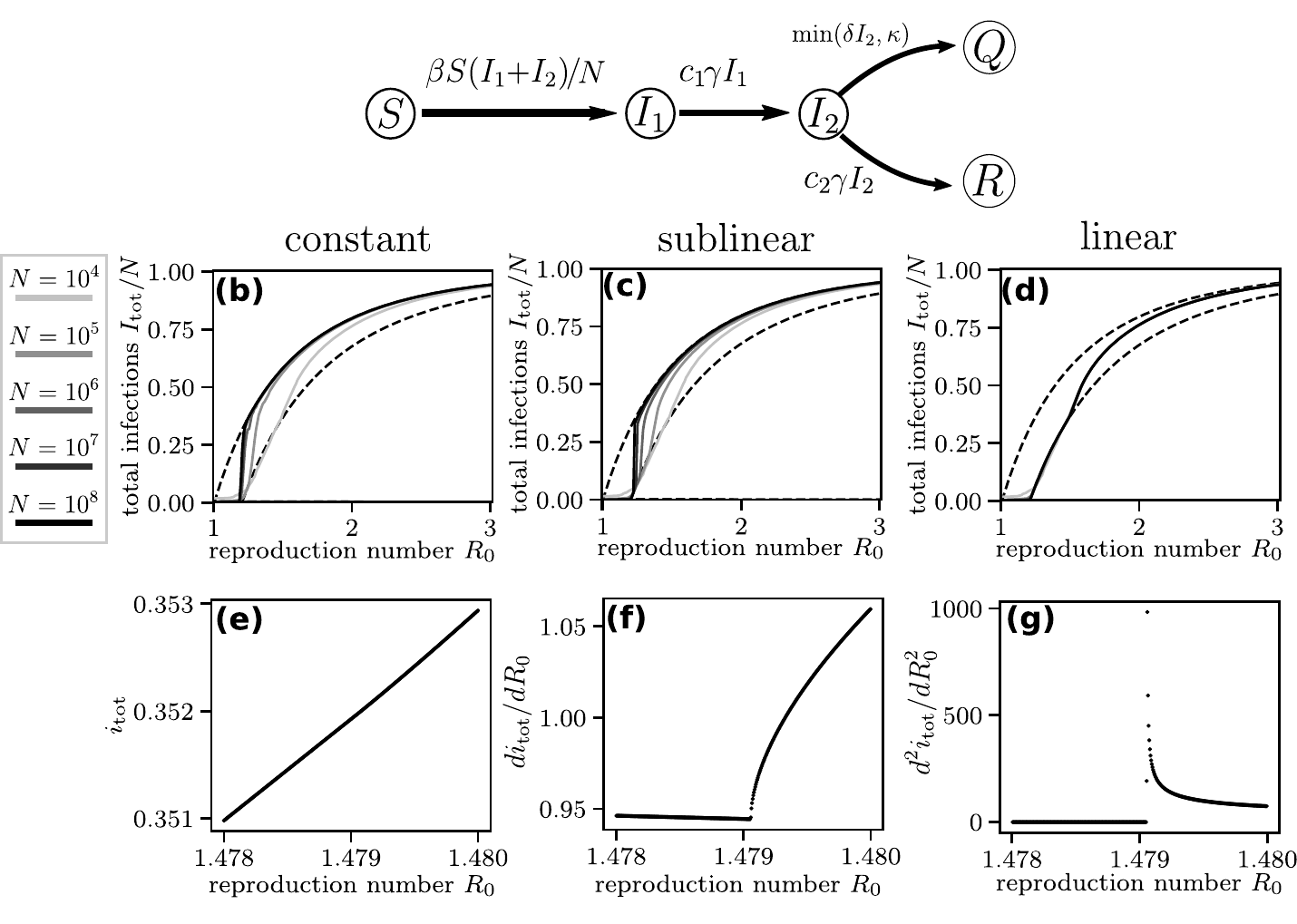}
        \caption{\textbf{Delayed effect of intervention measures.} 
        (a) Compartment model sketch. The model closely resembles the original SIRQ model analyzed in the main manuscript. However, the infected compartment is split into two compartments $I_1$ and $I_2$, with advance/recovery rates $c_1 \gamma$ and $c_2 \gamma$, respectively.
        (b-d) Average outbreak size $I_\mathrm{tot}/{N}$ as a function of the basic reproduction number $R_0$ for intervention capacity  $\kappa = 100$, $\kappa = 1\sqrt{N}$, and $\kappa = 0.01 N$ (identical for $N = 10^4$ for all three scaling regimes). Dashed lines present standard SIR dynamics with recovery rates $R_\mathrm{c}^{(1)} = 1$ and $R_\mathrm{c}^{(2)} = 6/5$, respectively. 
        All outbreak sizes are averaged only over large outbreaks $I_\mathrm{tot} > \sqrt{N}$ across a total of $100$ realizations with $I_1(0) = 10$ initially infected. 
        (e-g) Average outbreak size $i_\mathrm{tot}(R_0) = \lim_{N\rightarrow\infty} I_\mathrm{tot}(R_0) / N$ in the thermodynamic limit and its derivatives computed from mean-field differential equations for linearly scaling intervention capacity $\kappa = 0.01\,N$ with a fraction of $i_1(0) = 10^{-6}$ initial infected. As in the original SIRQ model, the second derivative $d^2i_\mathrm{tot}/dR_0^2$ is discontinuous, signifying the qualitative change of dynamics in a third-order transition (compare Fig.~3 in the main manuscript). All results are illustrated for $\gamma = 1$, $\delta = 1$ and $c_1 = c_2 = 2$.
         }
    \label{fig:SIIRQ_sat}
\end{figure}

An extension of the SIRQ model of the main manuscript explicitly includes disease progression in various stages or a time delay between infection and the onset of possible mitigation measures by splitting the infected compartment into two compartments $I_1$ and $I_2$, see Fig.~\ref{fig:SIIRQ_sat}a. 
For linearly scaling intervention capacity $\kappa = \tilde{\kappa}\, N$, the macroscopic dynamics are described by the mean-field deterministic rate equations
\begin{align*}
    \dfrac{ds}{dt} & = - \beta s (i_1+i_2) \\
    \dfrac{di_1}{dt} & = \beta s (i_1+i_2) - c_1 \gamma i_1   \\
      \dfrac{di_2}{dt} & = c_1\gamma i_1 - c_2\gamma i_2 -   \min[\delta i_2, \tilde{\kappa}]     \\
    \dfrac{dq}{dt} & =   \min[\delta i_2, \tilde{\kappa}] \\
    \dfrac{dr}{dt}  & = c_2 \gamma i_2 \,.
\end{align*}
Here, $c_1$ and $c_2$ denote the progression rate of the disease relative to the total recovery rate and are chosen such that the total recovery rate is $\left(\frac{1}{c_1\,\gamma} + \frac{1}{c_2\,\gamma}\right)^{-1} = \gamma$. The macroscopic dynamics of this model in the absence of interventions are thus equivalent to a standard SIR model.

With infinite intervention capacity, the effective recovery rate is given by $\gamma_\mathrm{eff} = \left(\frac{1}{c_1\,\gamma} + \frac{1}{c_2\,\gamma + \delta}\right)^{-1}$. For the parameters $\gamma = 1$, $c_1 = c_2 = 2$ and $\delta = 1$ this results in $\gamma_\mathrm{eff} = 6/5$ and consequently a critical point $R_\mathrm{c}^{(2)} = 6/5$. Since the intervention only acts after a delay and only leads to a shorter recovery time from $I_2$ to $R$, their effect in delaying the onset of macroscopic outbreaks is smaller than in the basic model studied in the main manuscript.

Figure~\ref{fig:SIIRQ_sat}b-d illustrates the average outbreak size $I_\mathrm{tot}/N$ as a function of the basic reproduction number $R_0$ for constant ($\alpha = 0$), sublinear ($0 <\alpha < 1$), and linear ($\alpha = 1$) intervention capacities $\kappa \propto N^\alpha$. The results are qualitatively identical to the observations in the main manuscript and feature the same three types of explosive phase transitions in the different scaling regimes. Figure~\ref{fig:SIIRQ_sat}e-g illustrates the third-order phase transition with linearly scaling intervention capacity.

\clearpage

 \subsection*{Targeted interventions}
\begin{figure}[h]
    \centering
    \includegraphics{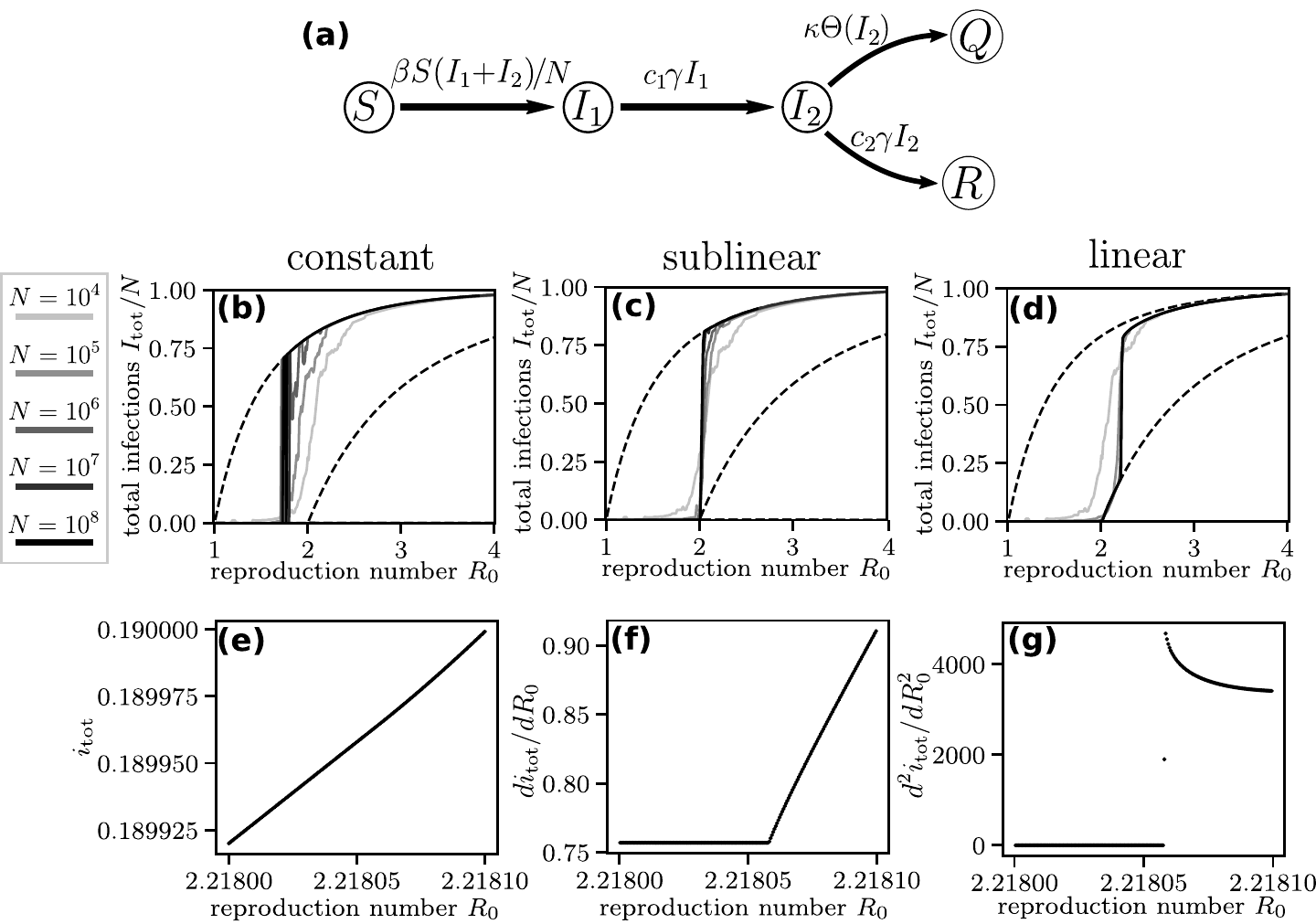}
        \caption{\textbf{Delayed, targeted interventions.} 
        (a) Compartment model sketch. Infected individuals are removed from $I_2$ with constant total rate $\kappa$. 
        (b-d) Average outbreak size $I_\mathrm{tot}/{N}$ as a function of the basic reproduction number $R_0$ for intervention capacity  $\kappa = 100$, $\kappa = 1\sqrt{N}$, and $\kappa = 0.01 N$ (identical for $N = 10^4$ for all three scaling regimes). Dashed lines present standard SIR dynamics with recovery rates $R_\mathrm{c}^{(1)} = 1$ and $R_\mathrm{c}^{(2)} = 2$, respectively.
        All outbreak sizes are averaged only over large outbreaks $I_\mathrm{tot} > \sqrt{N}$ across a total of $100$ realizations with $I_1(0) = 10$ initially infected. 
        (e-g) Average outbreak size $i_\mathrm{tot}(R_0) = \lim_{N\rightarrow\infty} I_\mathrm{tot}(R_0) / N$ in the thermodynamic limit and its derivatives computed from mean-field differential equations for linearly scaling intervention capacity $\kappa = 0.01\,N$ with a fraction of $i_1(0) = 10^{-6}$ initial infected. As in the original SIRQ model, the second derivative $d^2i_\mathrm{tot}/dR_0^2$ is discontinuous, signifying the qualitative change of dynamics in a third-order transition (compare Fig.~3 in the main manuscript). All results are illustrated for $\gamma = 1$ and $c_1 = c_2 = 2$.
        }
    \label{fig:SIIRQ_targeted}
\end{figure}

With the previous inclusion of delay, targeted interventions with a constant total rate $\kappa$ now also fulfill both conditions for explosive transitions (compare section I of this Supplemental Material). Figure~\ref{fig:SIIRQ_targeted}a shows the corresponding compartment model, the macroscopic dynamics with linearly scaling intervention capacity $\kappa = \tilde{\kappa}\, N$ are described by 
\begin{align*}
    \dfrac{ds}{dt} & = -  \beta s (i_1+i_2) \\
    \dfrac{di_1}{dt} & =  \beta s (i_1+i_2) - c_1 \gamma i_1   \\
      \dfrac{di_2}{dt} & = c_1\gamma i_1 - c_2\gamma i_2 -   \tilde{\kappa}\,\Theta\left(i_2\right)   \\
    \dfrac{dq}{dt} & = \tilde{\kappa}\,\Theta\left(i_2\right) \\
    \dfrac{dr}{dt}  & = c_2 \gamma i_2 \,,
\end{align*}
where $\Theta(\cdot)$ denotes the Heaviside step function and we again choose $\left(\frac{1}{c_1\,\gamma} + \frac{1}{c_2\,\gamma}\right)^{-1} = \gamma$.

With infinite intervention capacity, the infected are immediately removed from the second infected compartment $I_2$. The effective recovery rate is thus given by $\gamma_\mathrm{eff} = \left(\frac{1}{c_1\,\gamma}\right)^{-1} = c_1\,\gamma$. For the parameters $\gamma = 1$, $c_1 = c_2 = 2$ and $\delta = 1$ this results in $\gamma_\mathrm{eff} = 2$ and consequently a critical point $R_\mathrm{c}^{(2)} = 2$. Despite the infinite per-capita rate of the intervention compared to the basic model analyzed in the main manuscript, the delay prevents infected from being immediately removed and ensures that outbreaks do occur for sufficiently large $R_0$ even with infinite intervention capacity.

Figure~\ref{fig:SIIRQ_targeted}b-d illustrates the average outbreak size $I_\mathrm{tot}/N$ as a function of the basic reproduction number $R_0$ for constant ($\alpha = 0$), sublinear ($0 <\alpha < 1$), and linear ($\alpha = 1$) intervention capacities $\kappa \propto N^\alpha$. The results are qualitatively identical to the observations in the main manuscript and feature the same three types of explosive phase transitions in the different scaling regimes. Figure~\ref{fig:SIIRQ_targeted}e-g illustrates the third-order phase transition with linearly scaling intervention capacity.

\clearpage 

 \subsubsection*{More varied disease progression}
\begin{figure}[h]
    \centering
    \includegraphics{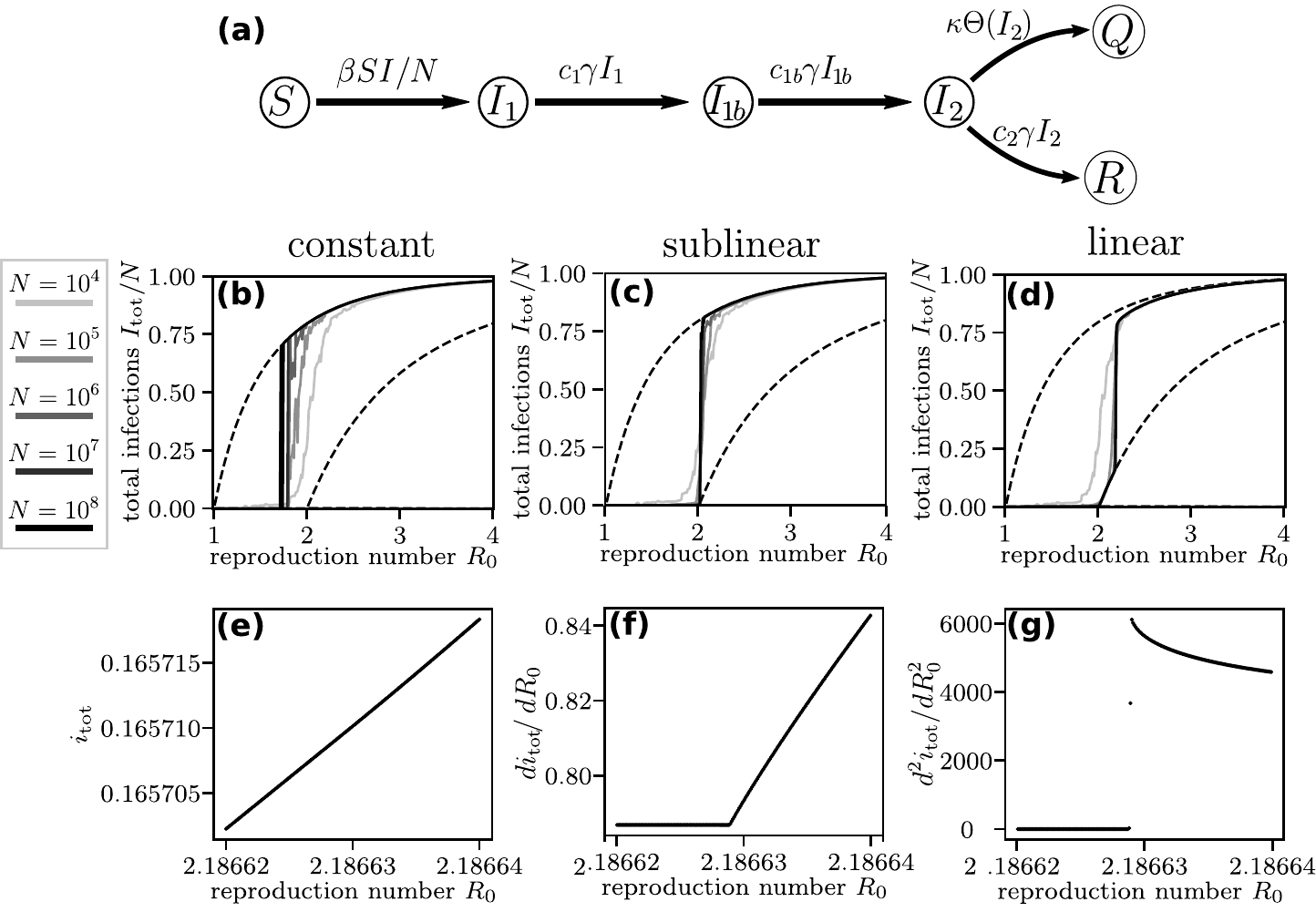}
        \caption{\textbf{Three stages of infection}. 
        (a) Compartment model sketch. Infected individuals move through successive stages of disease progression and are subject to targeted interventions with constant total rate $\kappa$ in the last stage $I_2$. 
        (b-d) Average outbreak size $I_\mathrm{tot}/{N}$ as a function of the basic reproduction number $R_0$ for intervention capacity  $\kappa = 100$, $\kappa = 1\sqrt{N}$, and $\kappa = 0.01 N$ (identical for $N = 10^4$ for all three scaling regimes). Dashed lines present standard SIR dynamics with recovery rates $R_\mathrm{c}^{(1)} = 1$ and $R_\mathrm{c}^{(2)} = 2$, respectively. 
        All outbreak sizes are averaged only over large outbreaks $I_\mathrm{tot} > \sqrt{N}$ across a total of $100$ realizations with $I_1(0) = 10$ initially infected. 
        (e-g) Average outbreak size $i_\mathrm{tot}(R_0) = \lim_{N\rightarrow\infty} I_\mathrm{tot}(R_0) / N$ in the thermodynamic limit and its derivatives computed from mean-field differential equations for linearly scaling intervention capacity $\kappa = 0.01\,N$ with a fraction of $i_1(0) = 10^{-6}$ initial infected. As in the original SIRQ model, the second derivative $d^2i_\mathrm{tot}/dR_0^2$ is discontinuous, signifying the qualitative change of dynamics in a third-order transition (compare Fig.~3 in the main manuscript). All results are illustrated for $\gamma = 1$, $c_1 = c_{1b} = 4$, and $c_2 = 2$.
        }
    \label{fig:SIIIRQ_linear}
\end{figure}

The details of the disease progression model do not affect our observations. In this model variation, we add an additional third infected compartment such that the states $I_1$, $I_{1b}$, and $I_2$ are passed successively, with targeted interventions with rate $\kappa$ only affecting individuals in the last one (Fig.~\ref{fig:SIIIRQ_linear}a). The macroscopic dynamics with linearly scaling intervention capacity $\kappa = \tilde{\kappa}\, N$ are described by
\begin{align*}
    \dfrac{ds}{dt} & = -  \beta s (i_1+i_{1b}+i_2) \\
    \dfrac{di_1}{dt} & =  \beta s (i_1+i_{1b}+i_2) - c_1 \gamma i_1   \\
    \dfrac{di_{1b}}{dt} & =  c_1 \gamma i_1 - c_{1b} \gamma i_{1b}   \\
    \dfrac{di_2}{dt} & = c_{1b} \gamma i_{1b} - c_2\gamma i_2 -   \tilde{\kappa}\,\Theta\left(i_2\right)   \\
    \dfrac{dq}{dt} & = \tilde{\kappa}\,\Theta\left(i_2\right) \\
    \dfrac{dr}{dt}  & = c_2 \gamma i_2 \,,
\end{align*}
where $\Theta(\cdot)$ denotes the Heaviside step function and we choose the recovery rate factors such that the total recovery rate without intervention is $\left(\frac{1}{c_1\,\gamma} + \frac{1}{c_{1b}\,\gamma} + \frac{1}{c_2\,\gamma}\right)^{-1} = \gamma$.

The macroscopic dynamics are, in fact, equivalent to the previous model. For infinite intervention capacity $\kappa = \infty$, the effective recovery rate is given by $\gamma_\mathrm{eff} = \left(\frac{1}{c_1\,\gamma} + \frac{1}{c_{1b}\,\gamma}\right)^{-1}$. For the parameters $\gamma = 1$, $c_1 = c_{1b} = 4$ and $c_2 = 2$ this results in $\gamma_\mathrm{eff} = 2$ and consequently a critical point $R_\mathrm{c}^{(2)} = 2$ as in the previous model.

Figure~\ref{fig:SIIIRQ_linear}b-d illustrate the average outbreak size $I_\mathrm{tot}/N$ as a function of the basic reproduction number $R_0$ for constant ($\alpha = 0$), sublinear ($0 <\alpha < 1$), and linear ($\alpha = 1$) intervention capacities $\kappa \propto N^\alpha$. The results are qualitatively identical to the observations in the main manuscript and feature the same three types of explosive phase transitions in the different scaling regimes. Figure~\ref{fig:SIIIRQ_linear}e-g illustrates the third-order phase transition with linearly scaling intervention capacity.

\clearpage 

 \subsection*{Partial interventions}

 \subsubsection*{Basic model variation}
\begin{figure}[h]
    \centering
    \includegraphics{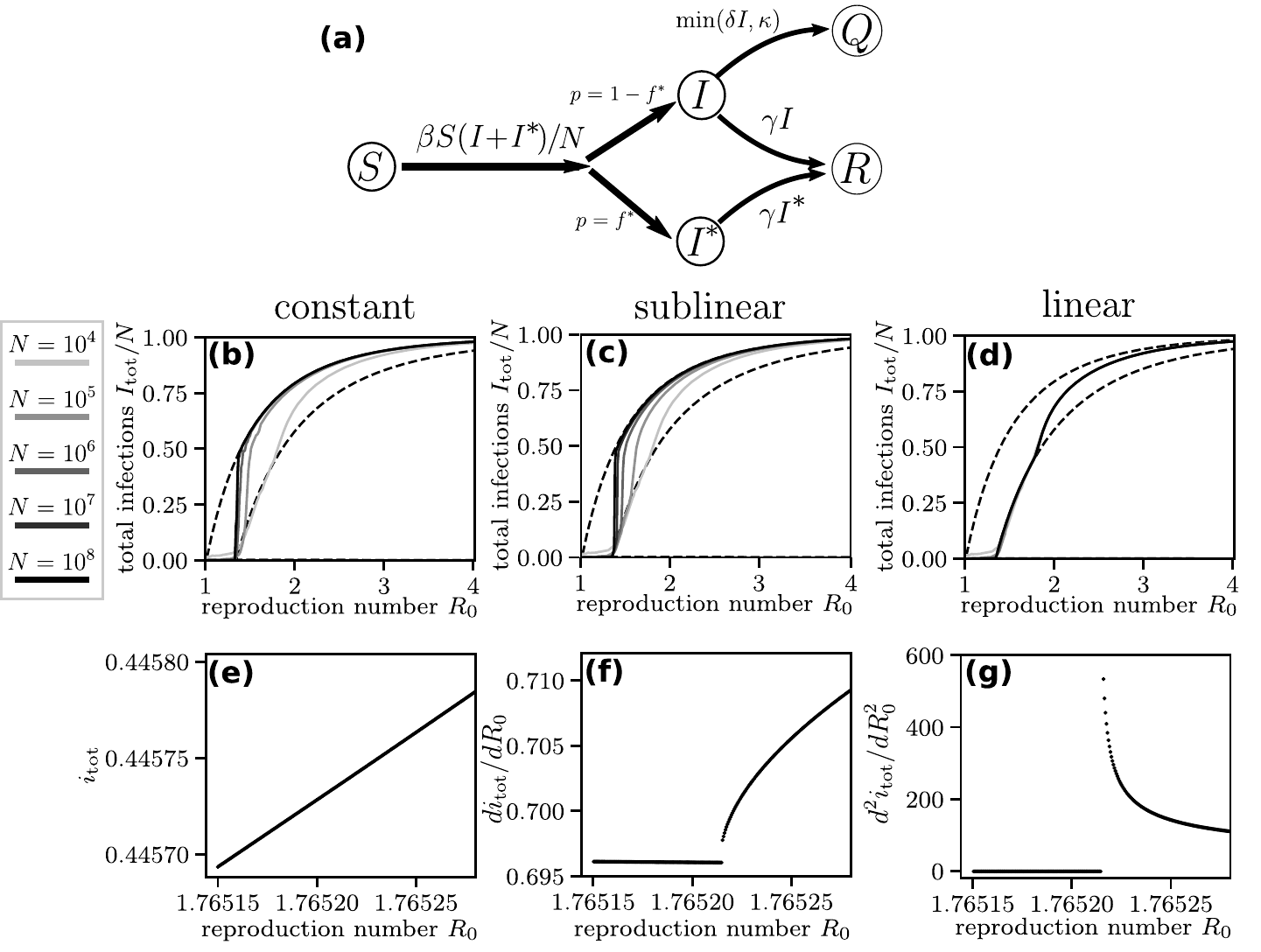}
    \caption{\textbf{Partial interventions.}
    (a) Compartment model sketch. A fraction $f^*$ of the population is not affected by the interventions, represented by the second, parallel infected compartment $I^*$.
    (b-d) Average outbreak size $I_\mathrm{tot}/{N}$ as a function of the basic reproduction number $R_0$ for intervention capacity  $\kappa = 100$, $\kappa = 1\sqrt{N}$, and $\kappa = 0.01 N$ (identical for $N = 10^4$ for all three scaling regimes). Dashed lines present standard SIR dynamics with recovery rates $R_\mathrm{c}^{(1)} = 1$ and $R_\mathrm{c}^{(2)} = 4/3$, respectively.
    All outbreak sizes are averaged only over large outbreaks $I_\mathrm{tot} > \sqrt{N}$ across a total of $100$ realizations with $I_1(0) = 10$ initially infected. 
    (e-g) Average outbreak size $i_\mathrm{tot}(R_0) = \lim_{N\rightarrow\infty} I_\mathrm{tot}(R_0) / N$ in the thermodynamic limit and its derivatives computed from mean-field differential equations for linearly scaling intervention capacity $\kappa = 0.01\,N$ with a fraction of $i_1(0) = 10^{-6}$ initial infected. As in the original SIRQ model, the second derivative $d^2i_\mathrm{tot}/dR_0^2$ is discontinuous, signifying the qualitative change of dynamics in a third-order transition (compare Fig.~3 in the main manuscript). All results are illustrated for $\gamma = 1$, $\delta = 1$, and $f^* = 1/2$.
    }
    \label{fig:SIIRQ_partial_testing}
\end{figure}

In this variation of the basic model discussed in the main manuscript, a fraction $f^*$ of the total population is not affected by the intervention, for example because they remain asymptomatic and never get tested. Figure~\ref{fig:SIIRQ_partial_testing}a shows the corresponding compartment model, the macroscopic dynamics with linearly scaling intervention capacity $\kappa = \tilde{\kappa}\, N$ are described by 
\begin{align*}
    \dfrac{ds}{dt} & = -  \beta s \left(i+i^*\right) \\
    \dfrac{di}{dt} & =  \left(1-f^*\right)\beta s \left(i+i^*\right) - \gamma i - \min[\delta i, \tilde{\kappa}] \\
      \dfrac{di^*}{dt} & = f^*\,\beta s \left(i+i^*\right) - \gamma i^*\\
    \dfrac{dq}{dt} & =   \min[\delta i, \tilde{\kappa}] \\
    \dfrac{dr}{dt}  & = \gamma\,\left(i+i^*\right) \,.
\end{align*}

With infinite intervention capacity, the expected time an infected individual remains infected is given by the weighted average of the effective recovery time. The fraction $f^*$ of infected that are not tested remain infected for the whole duration $1/\gamma$. The remaining infected are removed after an expected time $1/(\gamma + \delta)$. The effective recovery rate is thus given by $\gamma_\mathrm{eff} = \left(\frac{f^*}{\gamma} + \frac{1-f^*}{\gamma+\delta}\right)^{-1}$. For the parameters $\gamma = 1$, $f^* = 1/2$ and $\delta = 1$ this results in $\gamma_\mathrm{eff} = 4/3$ and consequently a critical point $R_\mathrm{c}^{(2)} = 4/3$. The effect of the intervention is naturally reduced compared to the basic model in the main manuscript where interventions affect the whole population.

Figure~\ref{fig:SIIRQ_partial_testing}b-d illustrates the average outbreak size $I_\mathrm{tot}/N$ as a function of the basic reproduction number $R_0$ for constant ($\alpha = 0$), sublinear ($0 <\alpha < 1$), and linear ($\alpha = 1$) intervention capacities $\kappa \propto N^\alpha$. The results are qualitatively identical to the observations in the main manuscript and feature the same three types of explosive phase transitions in the different scaling regimes. Figure~\ref{fig:SIIRQ_partial_testing}e-g illustrates the third-order phase transition with linearly scaling intervention capacity.

\clearpage

 \subsubsection*{Targeted interventions}
\begin{figure}[h]
    \centering
    \includegraphics{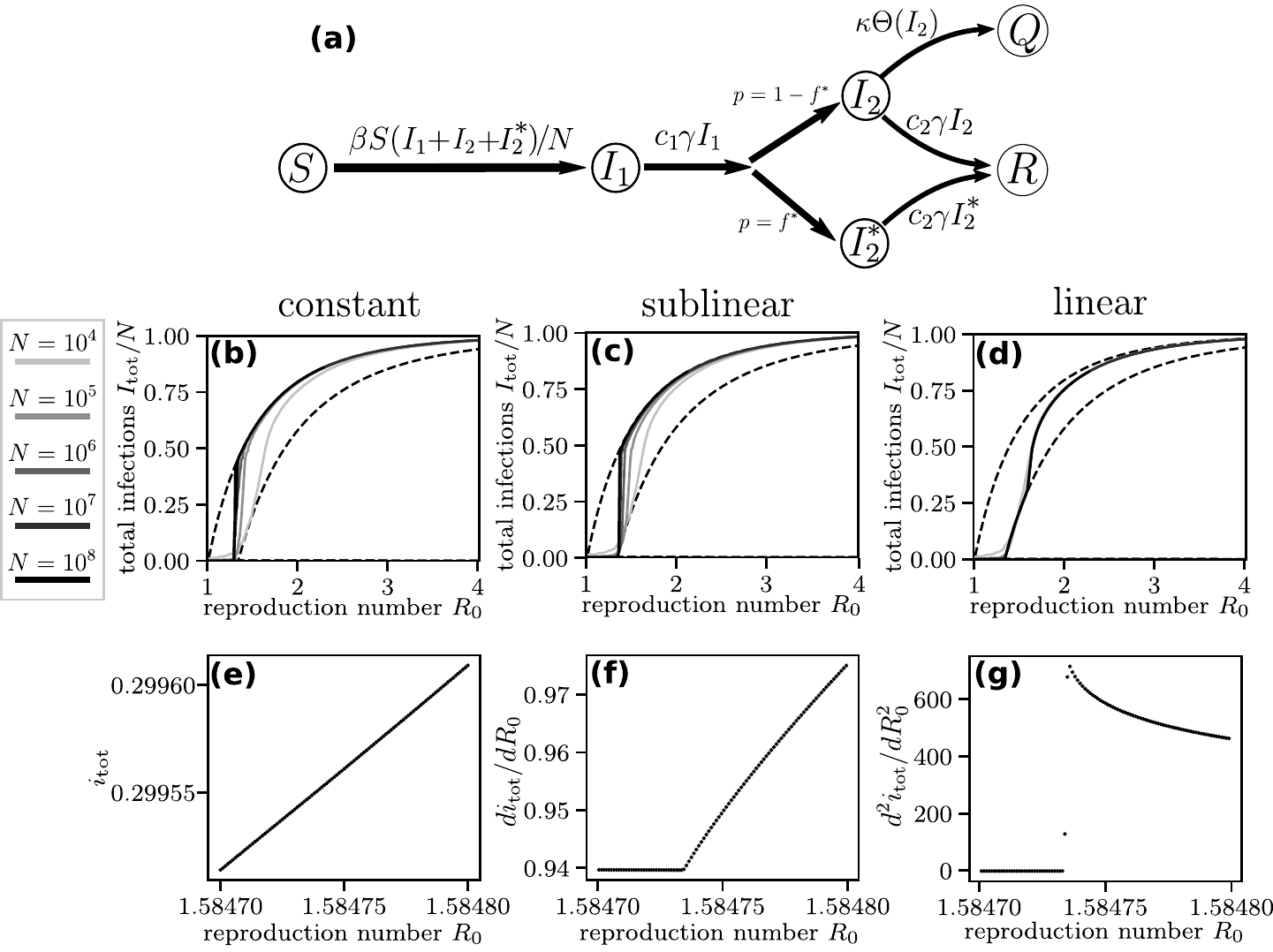}
    \caption{\textbf{Targeted partial interventions}. 
    (a) Compartment model sketch. Interventions are only effective after some delay when individuals enter $I_2$, but a fraction $f^*$ of the population is not affected by the interventions, represented by the second, parallel infected compartment $I_2^*$.
    (b-d) Average outbreak size $I_\mathrm{tot}/{N}$ as a function of the basic reproduction number $R_0$ for intervention capacity  $\kappa = 100$, $\kappa = 1\sqrt{N}$, and $\kappa = 0.01 N$ (identical for $N = 10^4$ for all three scaling regimes). Dashed lines present standard SIR dynamics with recovery rates $R_\mathrm{c}^{(1)} = 1$ and $R_\mathrm{c}^{(2)} = 4/3$, respectively.
    All outbreak sizes are averaged only over large outbreaks $I_\mathrm{tot} > \sqrt{N}$ across a total of $100$ realizations with $I_1(0) = 10$ initially infected. 
    (e-g) Average outbreak size $i_\mathrm{tot}(R_0) = \lim_{N\rightarrow\infty} I_\mathrm{tot}(R_0) / N$ in the thermodynamic limit and its derivatives computed from mean-field differential equations for linearly scaling intervention capacity $\kappa = 0.01\,N$ with a fraction of $i_1(0) = 10^{-6}$ initial infected. As in the original SIRQ model, the second derivative $d^2i_\mathrm{tot}/dR_0^2$ is discontinuous, signifying the qualitative change of dynamics in a third-order transition (compare Fig.~3 in the main manuscript). All results are illustrated for $\gamma = 1$, $c_1 = c_2 = 2$, and $f^* = 1/2$.
	}
    \label{fig:SIIIRQ_partial_testing}
\end{figure}

In this model variation, we consider targeted interventions that only affect a fraction $1-f^*$ of the total population. Figure~\ref{fig:SIIIRQ_partial_testing}a shows the corresponding compartment model, the macroscopic dynamics with linearly scaling intervention capacity $\kappa = \tilde{\kappa}\, N$ are described by 
\begin{align*}
    \dfrac{ds}{dt} & = -  \beta s \left(i_1 + i_2 + i_2^*\right) \\
    \dfrac{di_1}{dt} & =  \beta s \left(i_1 + i_2 + i_2^*\right) - c_1\gamma i_1 \\
    \dfrac{di_2}{dt} & = \left(1-f^*\right)c_1\gamma i_1 - c_2\gamma i_2 - \tilde{\kappa}\,\Theta\left(i_2\right) \\
    \dfrac{di_2^*}{dt} & = f^*c_1\gamma i_1 - c_2\gamma i_2^*\\
    \dfrac{dq}{dt} & =  \tilde{\kappa}\,\Theta\left(i_2\right) \\
    \dfrac{dr}{dt}  & = c_2\gamma\left(i_2+i_2^*\right) \,,
\end{align*}
where $\Theta(\cdot)$ denotes the Heaviside function. We again choose $c_1$ and $c_2$ such that the total recovery rate is $\left(\frac{1}{c_1\,\gamma} + \frac{1}{c_2\,\gamma}\right)^{-1} = \gamma$ and the macroscopic dynamics without interventions are equivalent to a standard SIR model.

With infinite intervention capacity, the expected time an infected individual remains infected is given by the weighted average of the effective recovery time. The fraction $f^*$ of infected that are not tested remain infected for the whole duration $1/\gamma$ while the remaining infected are immediately removed from state $I_2$ and only remain infected for a time $\frac{1}{c_1\gamma}$. The effective recovery rate is thus given by $\gamma_\mathrm{eff} = \left(\frac{f^*}{\gamma} + \frac{1-f^*}{c_1\gamma}\right)^{-1}$. For the parameters $\gamma = 1$, $f^* = 1/2$ and $c_1 = c_2 = 2$, this results in $\gamma_\mathrm{eff} = 4/3$, and consequently a critical point $R_\mathrm{c}^{(2)} = 4/3$.

Note that the delay in the intervention (i.e. the state $I_1$) is not strictly necessary as the model would also show an outbreak with immediate targeted interventions since a fraction of individuals is always infectious.

Figure~\ref{fig:SIIIRQ_partial_testing}b-d illustrates the average outbreak size $I_\mathrm{tot}/N$ as a function of the basic reproduction number $R_0$ for constant ($\alpha = 0$), sublinear ($0 <\alpha < 1$), and linear ($\alpha = 1$) intervention capacities $\kappa \propto N^\alpha$. The results are qualitatively identical to the observations in the main manuscript and feature the same three types of explosive phase transitions in the different scaling regimes. Figure~\ref{fig:SIIIRQ_partial_testing}e-g illustrates the third-order phase transition with linearly scaling intervention capacity.

\clearpage

 \subsection*{Ineffective interventions}

 \subsubsection*{Basic model variation}

\begin{figure}[h]
    \centering
    \includegraphics{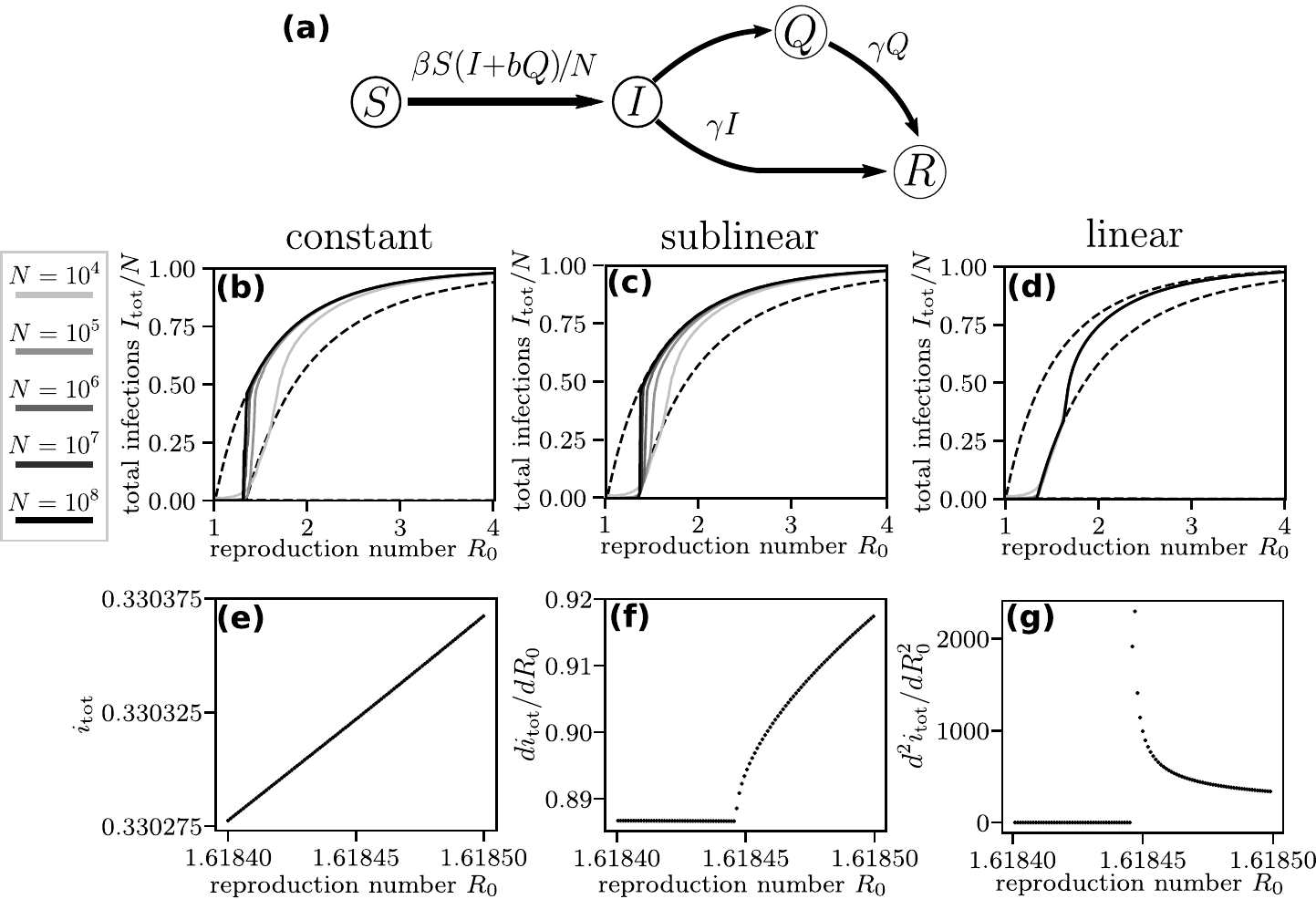}
        \caption{\textbf{Ineffective interventions.} 
        (a) Compartment model sketch. Interventions reduce the infection rate by a factor $b_Q$ but individuals still have to recover normally.
        (b-d) Average outbreak size $I_\mathrm{tot}/{N}$ as a function of the basic reproduction number $R_0$ for intervention capacity $\kappa = 100$, $\kappa = 1\sqrt{N}$, and $\kappa = 0.01 N$ (identical for $N = 10^4$ for all three scaling regimes). Dashed lines present standard SIR dynamics with recovery rates $R_\mathrm{c}^{(1)} = 1$ and $R_\mathrm{c}^{(2)} = 4/3$, respectively.
        All outbreak sizes are averaged only over large outbreaks $I_\mathrm{tot} > \sqrt{N}$ across a total of $100$ realizations with $I_1(0) = 10$ initially infected. 
        (e-g) Average outbreak size $i_\mathrm{tot}(R_0) = \lim_{N\rightarrow\infty} I_\mathrm{tot}(R_0) / N$ in the thermodynamic limit and its derivatives computed from mean-field differential equations for linearly scaling intervention capacity $\kappa = 0.01\,N$ with a fraction of $i_1(0) = 10^{-6}$ initial infected. As in the original SIRQ model, the second derivative $d^2i_\mathrm{tot}/dR_0^2$ is discontinuous, signifying the qualitative change of dynamics in a third-order transition (compare Fig.~3 in the main manuscript). All results are illustrated for $\gamma = 1$, $\delta = 1$, and $b_Q = 1/2$.
    }
    \label{fig:SIRQR_imperfect}
\end{figure}

In this basic model variation, individuals remain infectious after the intervention, though with a reduced infection rate $b_Q\,\beta$ with $0 < b_Q < 1$. Figure~\ref{fig:SIRQR_imperfect}a shows the corresponding compartment model, the macroscopic dynamics with linearly scaling intervention capacity $\kappa = \tilde{\kappa}\, N$ are described by 
\begin{align*}
    \dfrac{ds}{dt} & = - \beta s \left(i + b_Q q\right) \\
    \dfrac{di}{dt} & =  \beta s \left(i + b_Q q\right) - \gamma i - \min[\delta i, \tilde{\kappa}]\\
    \dfrac{dq}{dt} & =  \min[\delta i, \tilde{\kappa}] - \gamma q\\
    \dfrac{dr}{dt} & = \gamma \left(i + q\right) \,.
\end{align*}

The effective dynamics with infinite intervention capacity can here be understood in terms of an effective reduced infection rate $\beta_\mathrm{eff}$ instead of an effective increased recovery rate. Infected individuals fully recover after an expected time $1/\gamma$, independent of whether they are affected by the intervention or not. They initially cause new infections with a rate $\beta$ for an expected time $\frac{1}{\gamma + \delta}$. Afterwards, they are either removed or affected by the intervention with probability $\frac{\delta}{\gamma + \delta}$ and cause new infections with a reduced rate $b_Q\beta$ for their remaining expected recovery time $1/\gamma$. Overall, the effective infection rate is given by $\beta_\mathrm{eff} = \frac{\beta}{\gamma + \delta} + \frac{\delta}{\gamma + \delta}\,\frac{b_Q\beta}{\gamma}$. For the parameters $\gamma = 1$, $\delta = 1$, and $b_Q = 1/2$, this results in $\beta_\mathrm{eff} = 3/4$ and consequently a critical point $R_\mathrm{c}^{(2)} = 4/3$.

Figure~\ref{fig:SIRQR_imperfect}b-d illustrates the average outbreak size $I_\mathrm{tot}/N$ as a function of the basic reproduction number $R_0$ for constant ($\alpha = 0$), sublinear ($0 <\alpha < 1$), and linear ($\alpha = 1$) intervention capacities $\kappa \propto N^\alpha$. The results are qualitatively identical to the observations in the main manuscript and feature the same three types of explosive phase transitions in the different scaling regimes. Figure~\ref{fig:SIRQR_imperfect}e-g illustrates the third-order phase transition with linearly scaling intervention capacity.

\clearpage 

 \subsubsection*{Targeted interventions}
In this model variation, we consider targeted interventions that only reduce the infection rate of affected individuals. Figure~\ref{fig:SIIIRQ_partial_testing}a shows the corresponding compartment model, the macroscopic dynamics with linearly scaling intervention capacity $\kappa = \tilde{\kappa}\, N$ are described by 
\begin{align*}
    \dfrac{ds}{dt} & = - \beta s \left(i_1 + i_2 + b_Q q\right) \\
    \dfrac{di_1}{dt} & =  \beta s \left(i_1 + i_2 + b_Q q\right) - c_1 \gamma i_1\\
    \dfrac{di_2}{dt} & =  c_1 \gamma i_1 - c_2 \gamma i_2 - \tilde{\kappa} \Theta(i_2) \\
    \dfrac{dq}{dt} & = \tilde{\kappa} \Theta(i_2) - c_2\gamma q\\
    \dfrac{dr}{dt} & = c_2 \gamma \left(i_2 + q\right) \,.
\end{align*}
Note that the delay in the intervention is not strictly necessary here as the model would also show an outbreak with immediate targeted interventions as individuals in state $Q$ still cause new infections.

We again compute the effective infection rate $\beta_\mathrm{eff}$ in the limit of infinite intervention capacity. Infected individuals cause new infections with a rate $\beta$ before they advance to $I_2$ after an expected time $1/(c_1\gamma)$. They are then immediately affected by the intervention and moved to state $Q$, causing new infections with reduced rate $b_Q\beta$ for their remaining expected recovery time $1/(c_2\gamma)$. Overall, the effective infection rate is thus given by $\beta_\mathrm{eff} = \frac{\beta}{c_1\gamma} + \frac{b_Q\beta}{c_2\gamma}$. For the parameters $\gamma = 1$, $c_1 = c_2 = 2$ and $b_Q = 1/2$, this results in $\beta_\mathrm{eff} = 3/4$ and consequently a critical point $R_\mathrm{c}^{(2)} = 4/3$.

Figure~\ref{fig:SIRQR_imperfect}b-d illustrate the average outbreak size $I_\mathrm{tot}/N$ as a function of the basic reproduction number $R_0$ for constant ($\alpha = 0$), sublinear ($0 <\alpha < 1$), and linear ($\alpha = 1$) intervention capacities $\kappa \propto N^\alpha$. The results are qualitatively identical to the observations in the main manuscript and feature the same three types of explosive phase transitions in the different scaling regimes. Figure~\ref{fig:SIRQR_imperfect}e-g illustrates the third-order phase transition with linearly scaling intervention capacity.

\begin{figure}[h]
    \centering
    \includegraphics{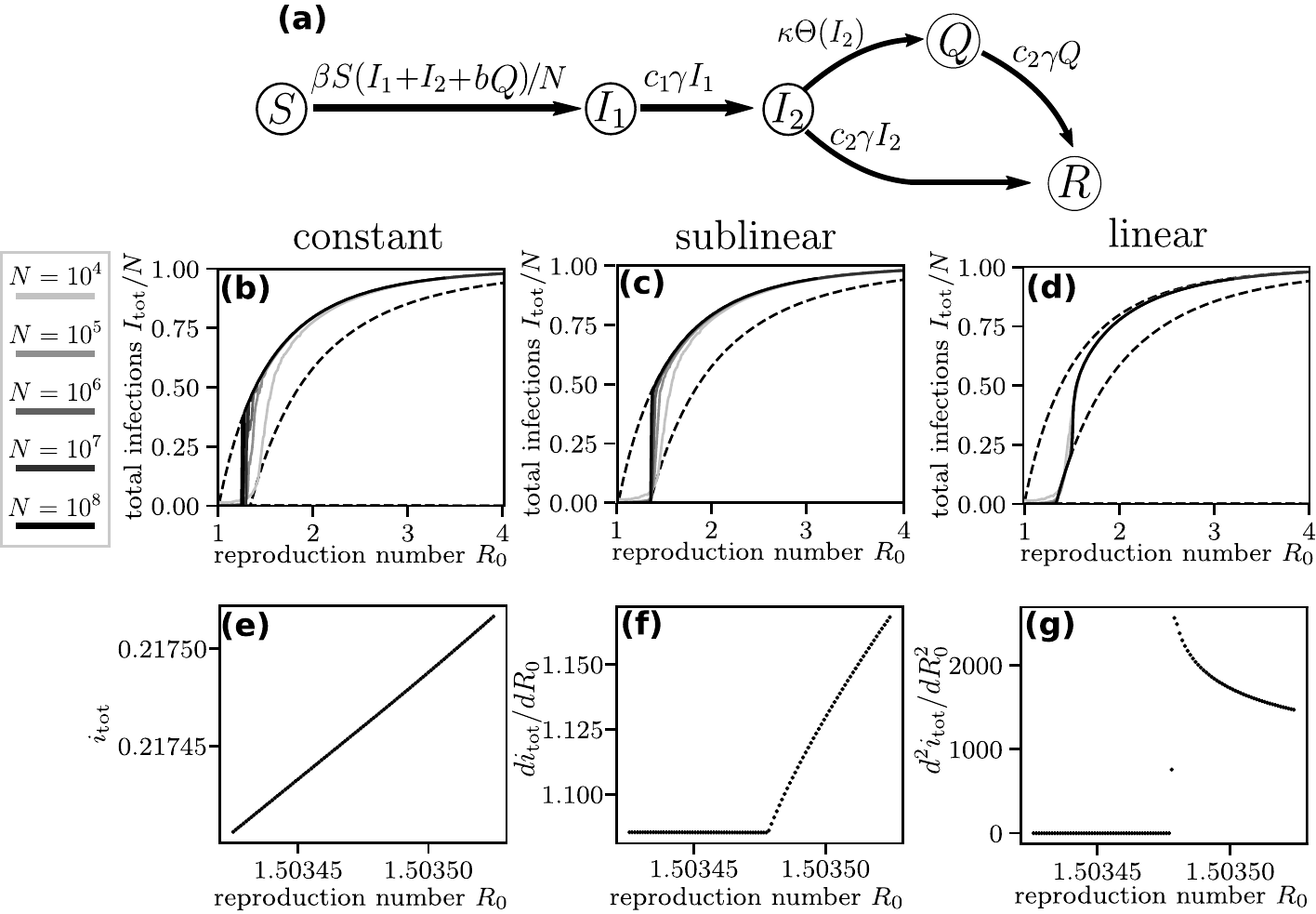}
        \caption{\textbf{Ineffective targeted interventions.} 
        (a) Compartment model sketch. Targeted interventions with a total rate $\kappa$ after a delay reduce the infection rate by a factor $b_Q$ but individuals still have to recover normally.
        (b-d) Average outbreak size $I_\mathrm{tot}/{N}$ as a function of the basic reproduction number $R_0$ for intervention capacity $\kappa = 100$, $\kappa = 1\sqrt{N}$, and $\kappa = 0.01 N$ (identical for $N = 10^4$ for all three scaling regimes). Dashed lines present standard SIR dynamics with recovery rates $R_\mathrm{c}^{(1)} = 1$ and $R_\mathrm{c}^{(2)} = 4/3$, respectively.
        All outbreak sizes are averaged only over large outbreaks $I_\mathrm{tot} > \sqrt{N}$ across a total of $100$ realizations with $I_1(0) = 10$ initially infected. 
        (e-g) Average outbreak size $i_\mathrm{tot}(R_0) = \lim_{N\rightarrow\infty} I_\mathrm{tot}(R_0) / N$ in the thermodynamic limit and its derivatives computed from mean-field differential equations for linearly scaling intervention capacity $\kappa = 0.01\,N$ with a fraction of $i_1(0) = 10^{-6}$ initial infected. As in the original SIRQ model, the second derivative $d^2i_\mathrm{tot}/dR_0^2$ is discontinuous, signifying the qualitative change of dynamics in a third-order transition (compare Fig.~3 in the main manuscript). All results are illustrated for $\gamma = 1$, $c_1 = c_2 = 2$, and $b_Q = 1/2$.
    }
    \label{fig:SIIRQR_imperfect}
\end{figure}

\clearpage

 \subsection*{Explicit time-delay dynamics}
\begin{figure}[h]
    \centering
    \includegraphics{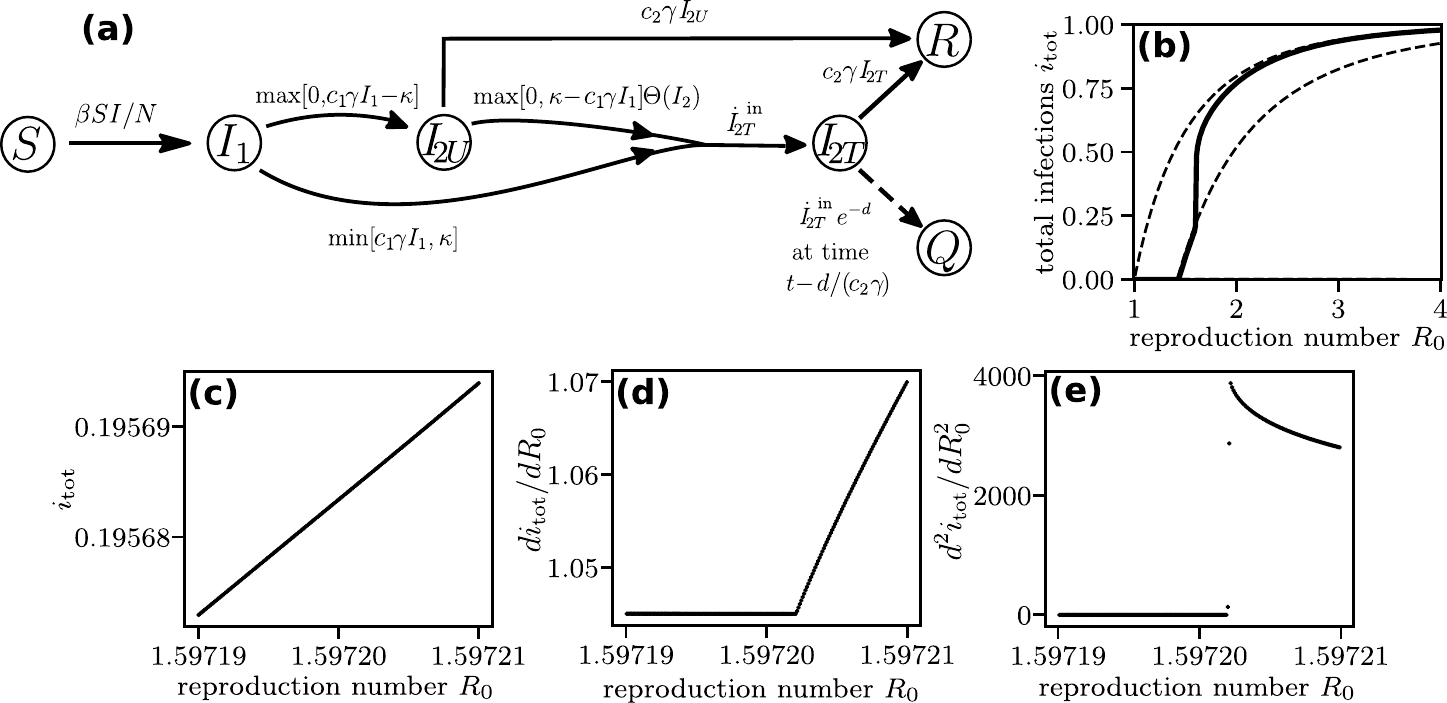}
        \caption{\textbf{Explicit time-delay dynamics.} 
        (a) Compartment model sketch illustrating the explicit time delay dynamics. When individuals would advance from $I_1$, they are either tested if sufficient capacity is available and move to $I_{2T}$ or otherwise remain untested and advance to $I_{2U}$, waiting to be tested or to recover. Individuals in $I_{2T}$ leave the system after a fixed delay $d/(c_2\gamma)$ but may naturally recover earlier. 
        (b) Average outbreak size $i_\mathrm{tot}(R_0) = \lim_{N\rightarrow\infty} I_\mathrm{tot}(R_0) / N$ in the thermodynamic limit as a function of the basic reproduction number $R_0$ from mean-field differential equations for linearly scaling intervention capacity $\kappa = 0.01\,N$ with a fraction of $i_1(0) = 10^{-6}$ initial infected. Dashed lines present standard SIR dynamics with recovery rates $R_\mathrm{c}^{(1)} = 1$ and $R_\mathrm{c}^{(2)} \approx 1.42$, respectively.
        (c-e) Average outbreak size $i_\mathrm{tot}(R_0)$ and its derivatives around the secondary transition. As in the original SIRQ model, the second derivative $d^2i_\mathrm{tot}/dR_0^2$ is discontinuous, signifying the qualitative change of dynamics in a third-order transition (compare Fig.~3 in the main manuscript). All results are illustrated for $\gamma = 1$, $c_1 = c_2 = 2$, and $d = 1/2$.
}
    \label{fig:SIIIRQ_delay}
\end{figure}

As a final model variation, we consider a more complex model with an explicit delay (Fig.~\ref{fig:SIIIRQ_delay}a). This may, for example, describe individuals waiting for tests results for a fixed time and isolating only afterwards to prevent further infections. The dynamics may be interpreted in terms of individuals noticing symptoms when they advance from state $I_1$. Individuals then immediately get tested if sufficient capacity is available and advance into $I_{2T}$. If the testing capacity is insufficient, they are not yet tested and advance into $I_{2U}$, where they wait for testing or natural recovery. Tested individuals in $I_{2T}$ recover naturally with rate $c_2 \gamma$. After a fixed time delay $d/(c_2 \gamma)$, the remaining fraction $e^{-d}$ of infected leaves the system due to the intervention into state $Q$. The macroscopic dynamics with linear intervention capacities $\kappa = \tilde{\kappa}/N$ are described by the corresponding mean-field delay differential equations
\begin{align*}
    \dfrac{ds}{dt} & = -  \beta s (i_1+i_{2U} +i_{2T}) \\
    \dfrac{di_1}{dt} & =  \beta s (i_1+i_{2U} +i_{2T}) - \max[0, c_1\gamma i_1 - \tilde{\kappa}] - \min [c_1\gamma i_1, \tilde{\kappa}]  \\
      \dfrac{di_{2U}}{dt} & = \max[0, c_1\gamma i_1 - \tilde{\kappa}] - \max[0,  \tilde{\kappa} - c_1\gamma i_1]\Theta(i_{2U}) - c_2 \gamma i_{2U}      \\
      \dfrac{di_{2T}}{dt} & =  \underbrace{\max[0,  \tilde{\kappa} - c_1\gamma i_1]\Theta(i_{2U}) + \min [c_1\gamma i_1, \tilde{\kappa}]}_{\dfrac{di_{2T}^\mathrm{in}}{dt}} - c_2\gamma i_{2T} - \left.\dfrac{di_{2T}^\mathrm{in}}{dt}\right|_{t-d/(c_2\gamma)}    \\
    \dfrac{dq}{dt} & = \left.\dfrac{di_{2T}^\mathrm{in}}{dt}\right|_{t-d/(c_2\gamma)}\\
    \dfrac{dr}{dt} & = c_2 \gamma i_{2U} + c_2 \gamma i_{2T} .
\end{align*}
Due to the increased complexity of the simulation, requiring to track individual infected to accurately model the microscopic dynamics with a fixed time delay, we focus on the solution of the delay differential equation and the dynamics with linear intervention capacity in the limit of infinite population only. Figure~\ref{fig:SIIIRQ_delay}b-d illustrates the emergence of the secondary third-order phase transition also in this model.

\newpage

\bibliography{}